\pdfoutput=1
\documentclass[11pt]{article}

\usepackage[final]{acl}

\usepackage{times}
\usepackage{latexsym}
\usepackage[T1]{fontenc}
\usepackage[utf8]{inputenc}
\usepackage{microtype}
\usepackage{xurl}

\usepackage{graphicx}
\usepackage{subcaption}
\usepackage{booktabs}
\usepackage{multirow}
\usepackage{enumitem}
\usepackage[table]{xcolor}
\usepackage{colortbl}
\usepackage{arydshln}
\usepackage{fancyvrb}
\usepackage{amsmath}
\usepackage{amssymb}
\usepackage{mathtools}
\usepackage[capitalize,noabbrev]{cleveref}
\usepackage{pgf}
\usepackage{placeins}
\usepackage{float}

\makeatletter
\ifacl@anonymize
  \newcommand{\codeurl}{\url{https://anonymous.4open.science/r/C2S-F530}}
\else
  \newcommand{\codeurl}{\url{https://github.com/dukesun99/Corpus2Skill}}
\fi
\makeatother

\newcommand{\corpusskill}{\textsc{Corpus2Skill}}
\newcommand{\skillmd}{\texttt{SKILL.md}}
\newcommand{\indexmd}{\texttt{INDEX.md}}

\definecolor{heathigh}{HTML}{C5E0B4}
\definecolor{costhigh}{HTML}{F4CCCC}
\definecolor{oursrow}{HTML}{E8F0FE}
\newcommand{\hc}[1]{%
  \pgfmathparse{int(min(100,max(0,#1*130)))}%
  \edef\hctemp{heathigh!\pgfmathresult}%
  \expandafter\cellcolor\expandafter{\hctemp}#1}
\newcommand{\hcb}[1]{%
  \pgfmathparse{int(min(100,max(0,#1*130)))}%
  \edef\hctemp{heathigh!\pgfmathresult}%
  \expandafter\cellcolor\expandafter{\hctemp}\textbf{#1}}
\newcommand{\hcost}[1]{%
  \pgfmathparse{int(min(100,max(0,#1*400)))}%
  \edef\hctemp{costhigh!\pgfmathresult}%
  \expandafter\cellcolor\expandafter{\hctemp}#1}
\newcommand{\hcostb}[1]{%
  \pgfmathparse{int(min(100,max(0,#1*400)))}%
  \edef\hctemp{costhigh!\pgfmathresult}%
  \expandafter\cellcolor\expandafter{\hctemp}\textbf{#1}}
\newcommand{\hh}[1]{%
  \pgfmathparse{int(min(100,max(0,#1*200)))}%
  \edef\hctemp{costhigh!\pgfmathresult}%
  \expandafter\cellcolor\expandafter{\hctemp}#1}
\newcommand{\hhb}[1]{%
  \pgfmathparse{int(min(100,max(0,#1*200)))}%
  \edef\hctemp{costhigh!\pgfmathresult}%
  \expandafter\cellcolor\expandafter{\hctemp}\textbf{#1}}

\title{Corpus2Skill: Distilling Enterprise Knowledge into\\Navigable Agent Skills for QA and RAG}

\author{
  \textbf{Yiqun Sun} \quad \textbf{Pengfei Wei}\thanks{\,Corresponding author.} \quad \textbf{Lawrence B.\ Hsieh} \\
  Magellan Technology Research Institute (MTRI) \\
  \texttt{\{duke.sun, pengfei.wei, lawrence.hsieh\}@mtri.co.jp}
}

\begin{document}
\maketitle

\begin{abstract}
Retrieval-Augmented Generation (RAG) grounds LLM responses in external evidence but treats the model as a passive consumer of search results, with no view of how the corpus is organized or what it has not yet seen.
We present \corpusskill{}, a system-level retrieval architecture for bounded, structurally coherent corpora such as enterprise knowledge bases: an offline compiler distills the corpus into a hierarchical skill directory, and at serve time an LLM agent navigates it, drilling from a bird's-eye view through progressively finer summaries down to documents and backtracking when a branch is unproductive.
On an enterprise customer-support benchmark, \corpusskill{} improves both answer quality and grounding over single-shot dense, hybrid, hierarchical-retrieval, and agentic RAG baselines at a moderate cost tradeoff, and the lead persists under encoder-matched controls and paired significance tests.
An eleven-dataset study shows that corpus navigation is \emph{not} a universal replacement for retrieval: it significantly wins on five datasets, ties on three, and loses on three. It helps on single-domain corpora with a recoverable topical taxonomy, but flat retrieval remains preferable on open-domain factoid pools or homogeneous-tabular corpora that defeat top-level clustering. We characterize this scope distinction as a design guideline for knowledge-grounded systems. Code is available at \codeurl{}.
\end{abstract}

\section{Introduction}
\label{sec:intro}

\begin{figure}[t]
  \centering
  \includegraphics[width=0.90\columnwidth]{figures/Corpus2Skill_teaser.pdf}
  \caption{\textbf{Retrieve vs.\ Navigate.} Traditional RAG passively feeds fixed retrieved passages to the LLM. The \emph{navigation} paradigm instead exposes the corpus as a structured hierarchy that the agent actively explores, backtracks through, and drills into to locate evidence.}
  \label{fig:teaser}
  \vskip -0.15in
\end{figure}

Enterprise knowledge bases contain thousands of heterogeneous support articles, product guides, and policy documents \citep{gao2024ragsurvey}. Retrieval-Augmented Generation \citep[RAG;][]{lewis2020retrieval} grounds LLM responses in such corpora by embedding the query, fetching similar passages from a vector index, and conditioning the answer on the retrieved context. This is now the dominant paradigm for enterprise QA, customer support, and compliance workflows \citep{gao2024ragsurvey,singh2025agenticragsurvey}. Yet for queries that touch several topics in the knowledge base, \textbf{the LLM never sees the forest for the trees}: it receives the top-$k$ passages but has no view of what else is in the corpus, no way to tell whether the best evidence was even in the retrieval window, and no map to plan a more thorough search \citep{sun2026information,tang2025uncovering}.

Agentic RAG \citep{yao2023react,schick2023toolformer} partially closes this gap by letting the agent issue multiple search queries, but each query is still a shot in the dark because the agent must guess productive search terms with no view of the corpus structure. Hierarchical approaches such as RAPTOR \citep{sarthi2024raptor}, GraphRAG \citep{edge2024graphrag}, StructRAG \citep{li2025structrag}, and BookRAG \citep{wang2025bookrag} organize documents into trees or graphs with LLM-generated summaries, but the hierarchy is still consumed indirectly through embedding-based search at query time, and the model receives results without seeing the organizational structure that produced them.

Meanwhile, a parallel line of work on LLM agent skills \citep{wang2023voyager,sok2026agentic} has shown that agents perform complex tasks more effectively when given structured, reusable knowledge modules that they can selectively load and follow. Agent platforms now support filesystem-based skill packages\footnote{\url{https://docs.anthropic.com/en/docs/agents-and-tools/agent-skills}} that use \emph{progressive disclosure} (\Cref{sec:progressive_disclosure}) to keep context costs low: the agent sees only lightweight metadata at startup and loads full content on demand, so the hierarchy can grow with the corpus without hitting context-window limits. Because skills are plain directories of files, they require no special infrastructure, which makes them a natural substrate for corpus navigation: rather than searching for passages, one can \emph{distill} the corpus into a navigable skill hierarchy and let the agent explore it.

\Cref{fig:teaser} contrasts these two paradigms. We propose exactly this with \corpusskill{}: a one-time offline compilation transforms the corpus into a navigable skill tree, and at serve time the LLM agent browses the tree to locate evidence. The hierarchy itself, not an embedding index, becomes the primary interface between the model and the corpus, giving the agent a bird's-eye view it can reason over before committing to any document. \corpusskill{} is a \emph{system-level} retrieval architecture: the compiled hierarchy, its enrichments, and the navigating agent operate as one pipeline, so we both evaluate the complete system against strong retrieval baselines and decompose which components carry the gains. Our contributions are:
\begin{itemize}[nolistsep,left=2pt]
\item We identify corpus \emph{navigability} as an underexplored axis in RAG and recast retrieval as agentic file browsing over a compiled skill tree.
\item We present a compile-then-navigate framework that turns a raw corpus into a hierarchical skill directory requiring no embedding index or vector DB at serve time.
\item We provide a quality- and grounding-aligned evaluation on WixQA \citep{cohen2025wixqa} against five baselines (BM25, Dense, Hybrid, RAPTOR, Agentic), with ablations plus encoder-matched, component, compiler, and 50k-document scale controls that trace the gains to the navigable structure.
\item We characterize \emph{when} corpus navigation helps through a 10-subset generalization sweep on RAGBench \citep{friel2024ragbench}, which with WixQA gives 11 datasets: under per-dataset paired significance tests \corpusskill{} wins on 5, ties on 3, and loses on 3. The split tracks corpus structure, and flat retrieval remains preferable on open-domain or homogeneous-tabular corpora (HAGRID, TatQA, CUAD).
\end{itemize}

\section{Background: Skills in LLM Agents}
\label{sec:skills_background}

\paragraph{From tools to skills.}
Modern LLM agents act on external systems through \emph{tools}: atomic, stateless functions such as web search or database queries \citep{schick2023toolformer,yao2023react}. \emph{Agent skills} \citep{sok2026agentic} fill the gap between acting and knowing \emph{how}: reusable modules that package procedural knowledge or multi-step workflows into a form the agent can load and follow. \citet{sok2026agentic} formalize a skill as a tuple $S = (C, \pi, T, R)$ of an applicability condition, an execution policy, a termination condition, and a reusable interface, distinguishing it from one-off plans, episodic memories, and atomic tools.

\paragraph{Filesystem-based skills.}
The current standard skill representation, adopted by Anthropic's Skills API\footnote{\url{https://www.anthropic.com/engineering/equipping-agents-for-the-real-world-with-agent-skills}} and recent open agent platforms, is a \emph{filesystem-based package}: a directory containing a \texttt{SKILL.md} alongside optional scripts and data assets. This format is favored because it cleanly supports \emph{progressive disclosure}: \emph{metadata}, specifically each skill's name and one-line description (typically under 100 tokens), is exposed at startup, the full \texttt{SKILL.md} \emph{instructions} are read only when the agent selects the skill, and \emph{resources} stay on disk until accessed. \corpusskill{} builds directly on this representation, mapping the corpus onto a forest of such skill directories.\label{sec:progressive_disclosure}

\paragraph{From procedural to informational skills.}
In prior work, skills encode \emph{how to perform a task}: workflows, decision trees, or code patterns. \corpusskill{} repurposes the same mechanism to encode \emph{what a corpus contains}: each skill directory represents a topical partition with \skillmd{} and \indexmd{} files at increasing detail, and the agent follows the same progressive-disclosure workflow (description $\to$ \skillmd{} $\to$ \indexmd{} $\to$ document) but navigates an information hierarchy rather than executing procedural instructions. This shift from procedural to informational skills is the central design insight of \corpusskill{}.

\section{Related Work}
\label{sec:related}

\paragraph{Cluster-based and structure-aware retrieval.}
Scatter/Gather \citep{cutting1992scatter} pioneered cluster-based navigation \citep{liu2004cluster} but required human selection. \corpusskill{} delegates this to an LLM agent. RAPTOR \citep{sarthi2024raptor} recursively clusters and summarizes chunks into a tree retrieved by traversal or collapsed-tree vector search. We share the compile-time embed-cluster-summarize loop but materialize the hierarchy as navigable files and replace fixed traversal with agentic navigation. GraphRAG \citep{edge2024graphrag} and HiRAG \citep{huang2025hirag} build knowledge graphs with community-level summaries but rely on graph retrieval at query time. RAG \citep{lewis2020retrieval,karpukhin2020dpr,khattab2022colbertv2} and its structure-aware extensions, including Self-RAG \citep{asai2024selfrag}, IRCoT \citep{trivedi2023ircot}, StructRAG \citep{li2025structrag}, BookRAG \citep{wang2025bookrag}, A-RAG \citep{du2026arag}, SPD-RAG \citep{akay2026spdrag}, NaviRAG \citep{dai2026navirag}, and HCAG \citep{wu2026hcag}, restructure, rerank \citep{sun2026very}, or re-query retrieved evidence but still depend on retrieval or search infrastructure at serve time \citep{gao2024ragsurvey,singh2025agenticragsurvey}.

\paragraph{Tool-use agents and skill construction.}
Toolformer \citep{schick2023toolformer} and ReAct \citep{yao2023react} establish LLMs' ability to select tools. We extend that paradigm with read-only file-browsing and document-lookup tools over a static hierarchy. Voyager \citep{wang2023voyager}, SkillX \citep{skillx2026}, and EvoSkill \citep{evoskill2026} build skills from agent \emph{trajectories}. \corpusskill{} builds them from \emph{document corpora} via clustering and summarization, a fundamentally different input modality.

\paragraph{Positioning.}
\corpusskill{} is distinguished from prior active-navigation work (NaviRAG, HCAG) by its filesystem-based skill representation with progressive disclosure via \skillmd{}/\indexmd{} files, and by its reliance on standard code-execution tools rather than custom retrieval APIs (\Cref{app:positioning} provides a feature comparison).

\section{\corpusskill{} Framework}
\label{sec:method}

\begin{figure*}[t]
  \vskip -0.05in
  \begin{center}
    \centerline{\includegraphics[width=0.92\textwidth]{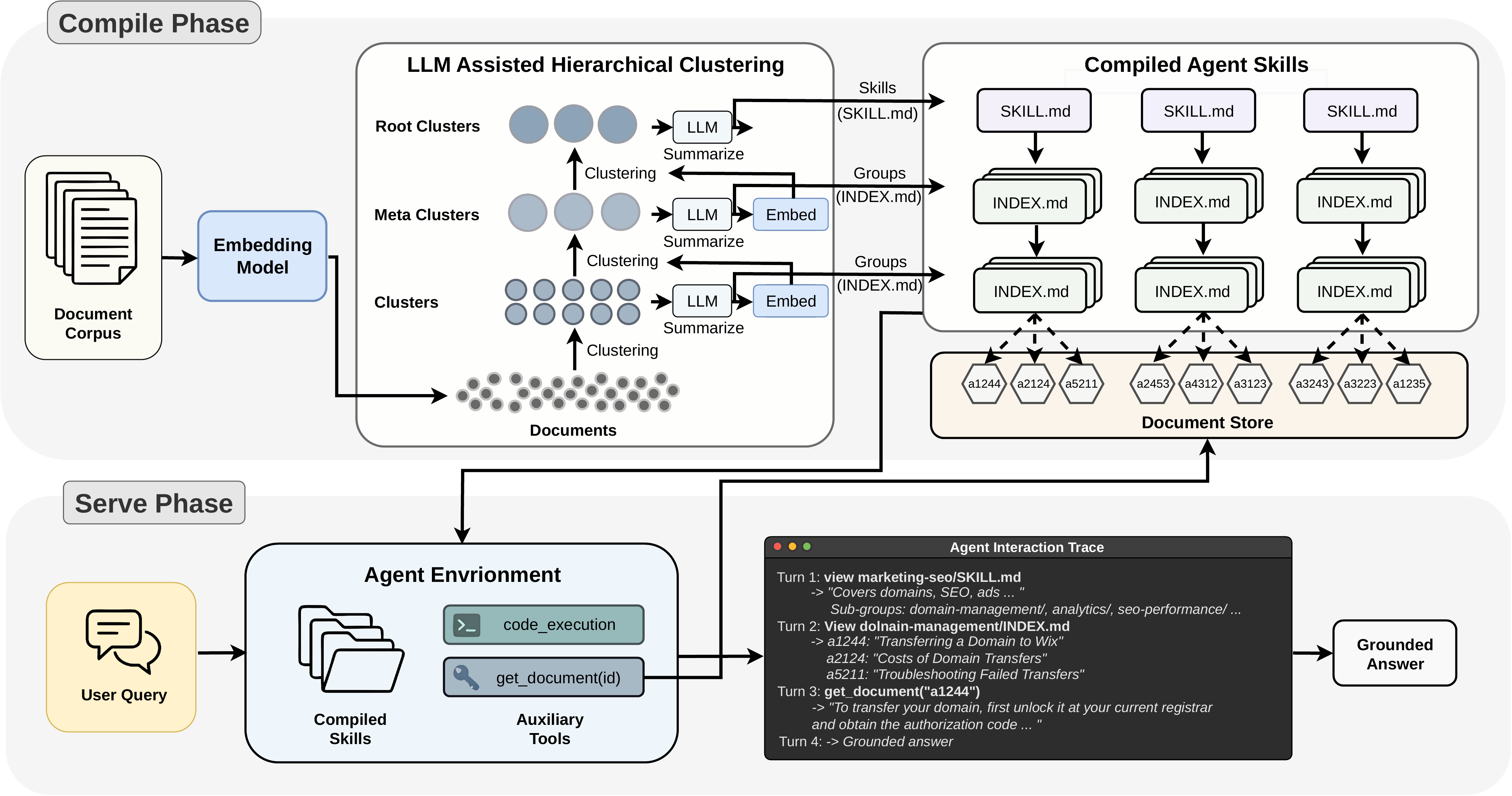}}
    \caption{\corpusskill{} system architecture. The compile phase (top) embeds documents, builds a multi-level cluster hierarchy with LLM summarization at each level, and materializes the result as a forest of navigable skills. The serve phase (bottom) uses an LLM agent with two tools: file navigation (\texttt{code\_execution}) and document lookup (\texttt{get\_document}).}
    \label{fig:system}
  \end{center}
  \vskip -0.25in
\end{figure*}

\corpusskill{} operates in two phases: a one-time offline \emph{compile phase} that transforms a corpus $\mathcal{D} = \{d_1, \dots, d_N\}$ into a hierarchical skill forest $\mathcal{S} = \{s_1, \dots, s_K\}$, and a per-query online \emph{serve phase} where an LLM agent navigates this tree with full visibility into the corpus structure, as shown in \Cref{fig:system}. Standard RAG selects $\mathcal{D}_q \subseteq \mathcal{D}$ via a black-box retrieval function $f(q, \mathcal{D}) \to \mathcal{D}_q$. \corpusskill{} replaces this with agent-driven navigation, where each skill $s_k$ is a navigable directory tree of summaries covering a topical partition of $\mathcal{D}$: the skill's summary serves as a routing condition, the agent descends through progressively finer summaries before retrieving documents, and the \skillmd{}/\indexmd{} files together with the \texttt{get\_document} tool form the reusable navigation interface.

\subsection{Compile Phase}
\label{sec:compile}

The compilation pipeline turns a raw corpus into a navigable forest of skills through a sequence of LLM-driven stages. Implementation details and full prompts are provided in \Cref{app:compilation,app:compile_prompts}.

\paragraph{Document representation.}
The compiler accepts standard document formats (\texttt{.md}, \texttt{.txt}, \texttt{.json}, \texttt{.jsonl}) and assigns each document a content-hash identifier. Every document is passed through a lightweight LLM call that returns a compact \emph{summary card} (title, one-line description, distinctive phrases). The card is concatenated with the truncated raw text and embedded with a sentence model \citep{reimers2019sentencebert}, so the vector reflects both literal content and an LLM-distilled gist, and the same cards reappear as \indexmd{} row entries.

\paragraph{Building the hierarchy.}
The hierarchy is built bottom-up under two parameters: the branching ratio $p$ and the maximum number of top-level clusters $K$. At each level the current $n$ embeddings are partitioned into $\lceil n/p \rceil$ groups via K-Means \citep{lloyd1982kmeans} on L2-normalized vectors. Each document is assigned a single home cluster, but documents whose top-1 versus top-2 cosine gap falls below a margin $\tau$ are additionally cross-referenced in the runner-up cluster's index, so the agent can discover cross-cutting content from multiple navigation paths. Each cluster is then summarized by an LLM, and a \emph{verify-and-repartition} pass examines all sibling summaries together to flag pairs that cover the same topic (\emph{confusable}) or single clusters that visibly span multiple topics (\emph{mixed}). Flagged clusters are repartitioned at the ID level and re-summarized. The cluster summaries, together with a small sample of member excerpts, are re-embedded and become the input to the next round of K-Means. The loop repeats until the number of remaining clusters falls below $K$. This embed-cluster-summarize recursion is inspired by RAPTOR \citep{sarthi2024raptor}, but materializes the hierarchy as browsable files rather than as a flat vector store.

\paragraph{Enabling cross-branch navigation.}
Each non-leaf node receives a short, filesystem-safe label (e.g., \texttt{wix-commerce-operations}) so directory names alone often identify the relevant branch. Once the hierarchy stabilizes, two enrichments are added. First, for every non-leaf cluster the top-$k$ documents closest to the centroid are surfaced as \emph{exemplars} in the navigation file, giving the agent concrete evidence of what each branch contains. Second, a batched LLM call tags every cluster summary with named entities and document types, aggregated into a corpus-wide \texttt{entity\_index.json}. Each skill's navigation file includes a \texttt{\#\# Related skills} block derived from entity-Jaccard overlap, so the agent can jump laterally to any other skill that mentions the same entity.

\paragraph{Skill tree construction.}
The hierarchy is materialized as a forest of $K$ skill directories: each root cluster becomes a \skillmd{} and each sub-cluster an \indexmd{} listing further subgroups or leaf document rows. Full document text lives separately in a \texttt{documents.json} accessed via \texttt{get\_document}, and a corpus-level \texttt{entity\_index.json} maps each named entity to the skill paths that mention it (full layout in \Cref{app:skill_structure}). This separation keeps navigation files small (typically under 2~KB each) so the agent can survey 20 documents per \indexmd{} cheaply before incurring the token cost of any full document.\footnote{Anthropic's current Skills API imposes limits of 8 skills per request, 200 files and 30\,MB per skill. These constraints informed our default parameters but are not fundamental to the framework design.} Each \skillmd{}/\indexmd{} follows a uniform template with YAML frontmatter (\texttt{name}, \texttt{description}, \texttt{level}, \texttt{num\_documents}) consumed by the Skills API for progressive disclosure, and the compilation prompts produce \emph{routing-oriented} summaries (topic area, question types, key terms) so that each file functions as a decision point for the navigating agent.

\paragraph{Navigation complexity.}
The hierarchy depth grows as $O(\log_p N)$, so the agent traverses at most $L = \lceil \log_p N \rceil$ levels and inspects $O(p \cdot \log_p N)$ summaries rather than $O(N)$. For WixQA ($N{=}6{,}221$, $p{=}10$), this gives $L{=}3$ and roughly 30 summaries per query; the framework therefore scales to $100$k-document corpora with one additional level (\Cref{app:complexity}).

\subsection{Serve Phase}
\label{sec:serve}

At serve time, the pre-compiled skill directories are uploaded to Anthropic's Skills API.
The API implements progressive disclosure: skill names and one-line descriptions are pre-loaded into the agent's context, but full file content is loaded only when the agent explicitly reads a file.
This design is critical for token efficiency.
Pre-loading all navigation files would flood the context with thousands of tokens of irrelevant summaries. Loading nothing would leave the agent blind to the corpus.
Progressive disclosure strikes a balance: the agent sees just enough (skill names and descriptions, approximately 200 tokens total) to make an informed first decision, then incrementally loads detail as it descends.

\paragraph{Tools.}
The agent receives two tools:
(1) \texttt{code\_execution}, which allows browsing \skillmd{} and \indexmd{} files (as well as the corpus-level \texttt{entity\_index.json}) through standard shell utilities such as \texttt{ls}, \texttt{cat}, and \texttt{grep}, providing the agent with full visibility into the hierarchy, and
(2) \texttt{get\_document(doc\_id)}, a custom tool that retrieves the full text of a document from the document store given its identifier.
The two-tool design reflects the two types of information the agent needs: \emph{navigational} (where to look) and \emph{evidential} (what to cite).
Keeping them separate means the agent can survey many document summaries cheaply before committing to the token cost of reading full documents.

\paragraph{Navigation workflow.}
Because skill descriptions are pre-loaded, the agent begins each query with a bird's-eye view of the corpus, knowing what topics exist and roughly how many documents each topic covers.
The system prompt instructs the agent to keep at least two candidate skills alive before committing to a path, to consult \texttt{entity\_index.json} when the query references a specific named entity, and to cross-jump via the \texttt{\#\# Related skills} block when its initial branch turns out to be thin.
A typical query proceeds in 2 to 3 turns:
\begin{enumerate}[nolistsep,left=2pt]
  \item The agent identifies the most relevant skill(s) from the pre-loaded descriptions (and from \texttt{entity\_index.json} if the query mentions a specific entity) and reads the corresponding \skillmd{}(s).
  \item The agent navigates into the most relevant subgroup's \indexmd{}, which lists document rows with titles, one-line descriptions, and distinctive phrases.
  \item The agent calls \texttt{get\_document} to retrieve the full text of one or more promising documents and synthesizes a grounded answer.
\end{enumerate}

At each step, the agent makes an informed decision guided by summaries describing each branch.
If a \skillmd{} lists ten subgroups and the agent has only explored one, it knows nine remain, an awareness that flat retrieval cannot provide.
This structural visibility enables two behaviors: \emph{targeted backtracking} (abandoning an unpromising branch for a more relevant one) and \emph{cross-branch synthesis} (combining evidence from multiple subgroups within or across skills), the latter aided by the related-skills cross-references and the entity index.
The system prompt enforces a \emph{grounding rule} (every claim must trace to a retrieved document, not a summary file) and a \emph{concise-answer format} (word limits, no preamble). See \Cref{app:system_prompt}.

\section{Experiments}
\label{sec:experiments}

\begin{table*}[t]
\centering
\small
\setlength{\tabcolsep}{4pt}
\renewcommand{\arraystretch}{1.08}
\caption{Main results on WixQA (200 queries). \corpusskill{} wins on F1, BERT, Fact., CtxR, CtxP and has the lowest Hallucination Rate among multi-turn methods; the F1 lead is bootstrap-significant ($p{<}0.0001$) with serve-time variance $\pm 0.002$ over three runs (\Cref{app:stats}). Single-shot retrievers win Faith.$_{\text{grd}}$ / Halluc.\ because they always feed back a small grounded passage. Best in \textbf{bold}; $\uparrow$ cells green (darker=better), $\downarrow$ cells (Halluc., \$/q) red (darker=worse). Multi-turn methods use Anthropic ephemeral prompt caching.}
\label{tab:main_results}
\begin{tabular}{l ccccccc rr}
\toprule
\multirow{2}{*}{\textbf{Method}} &
\multicolumn{7}{c}{\textbf{Quality Metrics}} &
\multicolumn{2}{c}{\textbf{Cost}} \\
\cmidrule(lr){2-8} \cmidrule(lr){9-10}
 & F1$\uparrow$ & BERT$\uparrow$ & Fact.$\uparrow$ & Faith.$_{\text{grd}}\uparrow$ & CtxR$\uparrow$ & CtxP$\uparrow$ & Halluc.$\downarrow$ & Turns & \$/q$\downarrow$ \\
\midrule
BM25
  & \hc{0.345} & \hc{0.830} & \hc{0.472} & \hc{0.831} & \hc{0.406} & \hc{0.549} & \hh{0.005}
  & 1.00 & \hcostb{0.007} \\
Dense
  & \hc{0.364} & \hc{0.833} & \hc{0.537} & \hc{0.889} & \hc{0.448} & \hc{0.659} & \hhb{0.000}
  & 1.00 & \hcost{0.007} \\
Hybrid
  & \hc{0.361} & \hc{0.832} & \hc{0.524} & \hc{0.851} & \hc{0.416} & \hc{0.637} & \hh{0.005}
  & 1.00 & \hcost{0.007} \\
RAPTOR
  & \hc{0.398} & \hc{0.839} & \hc{0.704} & \hcb{0.909} & \hc{0.618} & \hc{0.659} & \hh{0.005}
  & 1.00 & \hcost{0.012} \\
Agentic
  & \hc{0.378} & \hc{0.843} & \hc{0.705} & \hc{0.528} & \hc{0.452} & \hc{0.649} & \hh{0.500}
  & 5.06 & \hcost{0.082} \\
\midrule
\corpusskill{}
  & \hcb{0.456} & \hcb{0.862} & \hcb{0.767} & \hc{0.859} & \hcb{0.708} & \hcb{0.829} & \hh{0.045}
  & 2.38 & \hcost{0.153} \\
\bottomrule
\end{tabular}
\vspace{-0.5em}
\end{table*}

\begin{table*}[!t]
\centering
\scriptsize
\setlength{\tabcolsep}{2.5pt}
\renewcommand{\arraystretch}{0.98}
\caption{Generalization on 10 RAGBench subsets (200 queries each, or all available) over the \emph{expanded retrieval pool}, grouped by the structural pattern that determines whether navigation helps. $^{\dagger}$/$^{\ddagger}$ marks a significant \corpusskill{} win/loss in Token F1 vs.\ the best baseline (paired bootstrap, $p{<}0.05$; \Cref{app:stats}); unmarked \corpusskill{} rows are ties. Per-dataset best in \textbf{bold}; $\uparrow$ cells green, $\downarrow$ cells (Halluc., \$/q) red.}
\label{tab:ragbench_results}
\begin{tabular}{l l ccccccc rr}
\toprule
\multirow{2}{*}{\textbf{Dataset}} & \multirow{2}{*}{\textbf{Method}} &
\multicolumn{7}{c}{\textbf{Quality Metrics}} &
\multicolumn{2}{c}{\textbf{Cost}} \\
\cmidrule(lr){3-9} \cmidrule(lr){10-11}
 & & F1$\uparrow$ & BERT$\uparrow$ & Fact.$\uparrow$ & Faith.$_{\text{grd}}\uparrow$ & CtxR$\uparrow$ & CtxP$\uparrow$ & Halluc.$\downarrow$ & Turns & \$/q$\downarrow$ \\
\midrule
\multicolumn{11}{l}{\textit{(a) Single-domain, atomic-document corpora: \corpusskill{} leads (significantly on all but TechQA)}} \\
\midrule
\multirow{5}{*}{Emanual}
  & BM25     & \hc{0.412} & \hc{0.855} & \hc{0.661} & \hc{0.874} & \hc{0.556} & \hc{0.586} & \hhb{0.000} & 1.00 & \hcost{0.006} \\
  & Dense    & \hc{0.421} & \hc{0.855} & \hc{0.691} & \hc{0.894} & \hc{0.650} & \hc{0.668} & \hhb{0.000} & 1.00 & \hcost{0.006} \\
  & Hybrid   & \hc{0.424} & \hc{0.854} & \hc{0.721} & \hcb{0.895} & \hc{0.647} & \hc{0.644} & \hhb{0.000} & 1.00 & \hcost{0.006} \\
  & Agentic  & \hc{0.455} & \hc{0.871} & \hcb{0.841} & \hc{0.696} & \hc{0.611} & \hc{0.633} & \hh{0.220} & 3.39 & \hcost{0.046} \\
  & \corpusskill{}$^{\dagger}$ & \hcb{0.497} & \hcb{0.898} & \hc{0.817} & \hc{0.835} & \hcb{0.768} & \hcb{0.832} & \hh{0.091} & 2.57 & \hcost{0.060} \\
\midrule
\multirow{5}{*}{DelucionQA}
  & BM25     & \hc{0.412} & \hc{0.858} & \hc{0.730} & \hc{0.946} & \hc{0.681} & \hc{0.664} & \hh{0.005} & 1.00 & \hcost{0.006} \\
  & Dense    & \hc{0.440} & \hc{0.862} & \hc{0.821} & \hcb{0.969} & \hc{0.750} & \hc{0.753} & \hhb{0.000} & 1.00 & \hcost{0.006} \\
  & Hybrid   & \hc{0.429} & \hc{0.861} & \hc{0.775} & \hc{0.957} & \hc{0.714} & \hc{0.733} & \hhb{0.000} & 1.00 & \hcost{0.006} \\
  & Agentic  & \hc{0.438} & \hc{0.874} & \hcb{0.841} & \hc{0.719} & \hc{0.711} & \hc{0.725} & \hh{0.206} & 2.91 & \hcost{0.039} \\
  & \corpusskill{}$^{\dagger}$ & \hcb{0.478} & \hcb{0.893} & \hc{0.821} & \hc{0.921} & \hcb{0.824} & \hcb{0.872} & \hh{0.022} & 2.64 & \hcost{0.110} \\
\midrule
\multirow{5}{*}{TechQA}
  & BM25     & \hc{0.377} & \hc{0.840} & \hc{0.537} & \hcb{0.781} & \hc{0.295} & \hc{0.432} & \hh{0.040} & 1.00 & \hcostb{0.009} \\
  & Dense    & \hc{0.382} & \hc{0.840} & \hc{0.560} & \hc{0.774} & \hc{0.297} & \hc{0.499} & \hh{0.050} & 1.00 & \hcost{0.009} \\
  & Hybrid   & \hc{0.383} & \hc{0.840} & \hc{0.562} & \hc{0.779} & \hc{0.267} & \hc{0.466} & \hhb{0.040} & 1.00 & \hcost{0.009} \\
  & Agentic  & \hc{0.309} & \hc{0.836} & \hc{0.605} & \hc{0.414} & \hc{0.271} & \hc{0.476} & \hh{0.635} & 6.33 & \hcost{0.151} \\
  & \corpusskill{} & \hcb{0.396} & \hcb{0.854} & \hcb{0.709} & \hc{0.672} & \hcb{0.496} & \hcb{0.649} & \hh{0.280} & 3.75 & \hcost{0.507} \\
\midrule
\multirow{5}{*}{ExpertQA}
  & BM25     & \hc{0.312} & \hc{0.825} & \hc{0.486} & \hcb{0.913} & \hc{0.439} & \hc{0.408} & \hhb{0.005} & 1.00 & \hcostb{0.007} \\
  & Dense    & \hc{0.330} & \hc{0.827} & \hc{0.573} & \hc{0.909} & \hc{0.529} & \hc{0.507} & \hh{0.010} & 1.00 & \hcost{0.007} \\
  & Hybrid   & \hc{0.326} & \hc{0.827} & \hc{0.552} & \hc{0.898} & \hc{0.478} & \hc{0.463} & \hh{0.015} & 1.00 & \hcost{0.007} \\
  & Agentic  & \hc{0.288} & \hc{0.838} & \hc{0.709} & \hc{0.517} & \hc{0.442} & \hc{0.457} & \hh{0.508} & 4.39 & \hcost{0.085} \\
  & \corpusskill{}$^{\dagger}$ & \hcb{0.356} & \hcb{0.854} & \hcb{0.764} & \hc{0.666} & \hcb{0.565} & \hcb{0.589} & \hh{0.225} & 3.88 & \hcost{0.183} \\
\midrule
\multirow{5}{*}{ClapNQ}
  & BM25     & \hc{0.325} & \hc{0.849} & \hc{0.735} & \hc{0.940} & \hc{0.766} & \hc{0.406} & \hhb{0.000} & 1.00 & \hcostb{0.005} \\
  & Dense    & \hc{0.333} & \hc{0.850} & \hc{0.774} & \hc{0.934} & \hc{0.788} & \hc{0.420} & \hh{0.005} & 1.00 & \hcost{0.005} \\
  & Hybrid   & \hc{0.333} & \hc{0.850} & \hc{0.756} & \hc{0.938} & \hc{0.782} & \hc{0.411} & \hh{0.005} & 1.00 & \hcost{0.005} \\
  & Agentic  & \hc{0.308} & \hc{0.855} & \hc{0.756} & \hc{0.736} & \hc{0.710} & \hc{0.421} & \hh{0.161} & 2.52 & \hcost{0.027} \\
  & \corpusskill{}$^{\dagger}$ & \hcb{0.468} & \hcb{0.893} & \hcb{0.924} & \hcb{0.911} & \hcb{0.937} & \hcb{0.878} & \hh{0.015} & 2.03 & \hcost{0.080} \\
\midrule
\multicolumn{11}{l}{\textit{(b) Heterogeneous or numerical-table corpora: statistical ties}} \\
\midrule
\multirow{5}{*}{CovidQA}
  & BM25     & \hc{0.204} & \hc{0.785} & \hc{0.546} & \hc{0.905} & \hc{0.542} & \hc{0.517} & \hh{0.050} & 1.00 & \hcostb{0.006} \\
  & Dense    & \hc{0.215} & \hc{0.804} & \hc{0.635} & \hcb{0.927} & \hc{0.634} & \hcb{0.602} & \hhb{0.030} & 1.00 & \hcost{0.006} \\
  & Hybrid   & \hcb{0.216} & \hc{0.795} & \hc{0.613} & \hc{0.924} & \hc{0.608} & \hc{0.602} & \hh{0.035} & 1.00 & \hcost{0.006} \\
  & Agentic  & \hc{0.206} & \hcb{0.849} & \hcb{0.651} & \hc{0.612} & \hc{0.549} & \hc{0.617} & \hh{0.363} & 4.13 & \hcost{0.058} \\
  & \corpusskill{} & \hc{0.208} & \hc{0.847} & \hc{0.456} & \hc{0.735} & \hc{0.405} & \hc{0.426} & \hh{0.255} & 6.85 & \hcost{0.436} \\
\midrule
\multirow{5}{*}{FinQA}
  & BM25     & \hc{0.359} & \hc{0.847} & \hc{0.340} & \hcb{0.940} & \hc{0.358} & \hc{0.413} & \hh{0.015} & 1.00 & \hcostb{0.005} \\
  & Dense    & \hcb{0.400} & \hc{0.852} & \hc{0.520} & \hc{0.924} & \hc{0.514} & \hc{0.479} & \hh{0.020} & 1.00 & \hcost{0.006} \\
  & Hybrid   & \hc{0.389} & \hc{0.853} & \hc{0.440} & \hc{0.936} & \hc{0.438} & \hc{0.481} & \hhb{0.005} & 1.00 & \hcost{0.006} \\
  & Agentic  & \hc{0.368} & \hc{0.861} & \hcb{0.576} & \hc{0.618} & \hc{0.512} & \hc{0.506} & \hh{0.420} & 4.36 & \hcost{0.076} \\
  & \corpusskill{} & \hc{0.392} & \hcb{0.867} & \hc{0.574} & \hc{0.770} & \hcb{0.550} & \hcb{0.652} & \hh{0.175} & 4.51 & \hcost{0.495} \\
\midrule
\multicolumn{11}{l}{\textit{(c) Open-domain factoid or homogeneous-tabular corpora: \corpusskill{} loses}} \\
\midrule
\multirow{5}{*}{HAGRID}
  & BM25     & \hc{0.320} & \hc{0.862} & \hc{0.815} & \hc{0.967} & \hc{0.825} & \hc{0.613} & \hh{0.010} & 1.00 & \hcostb{0.005} \\
  & Dense    & \hc{0.322} & \hc{0.862} & \hc{0.855} & \hcb{0.974} & \hcb{0.881} & \hc{0.627} & \hhb{0.000} & 1.00 & \hcost{0.005} \\
  & Hybrid   & \hc{0.323} & \hc{0.862} & \hc{0.831} & \hc{0.970} & \hc{0.859} & \hcb{0.641} & \hh{0.005} & 1.00 & \hcost{0.005} \\
  & Agentic  & \hcb{0.356} & \hcb{0.879} & \hc{0.826} & \hc{0.881} & \hc{0.829} & \hc{0.623} & \hh{0.071} & 2.29 & \hcost{0.022} \\
  & \corpusskill{}$^{\ddagger}$ & \hc{0.188} & \hc{0.838} & \hc{0.358} & \hc{0.787} & \hc{0.220} & \hc{0.200} & \hh{0.260} & 1.05 & \hcost{0.021} \\
\midrule
\multirow{5}{*}{TatQA}
  & BM25     & \hc{0.307} & \hc{0.847} & \hc{0.468} & \hcb{0.965} & \hc{0.471} & \hc{0.489} & \hhb{0.005} & 1.00 & \hcostb{0.005} \\
  & Dense    & \hc{0.312} & \hc{0.846} & \hc{0.527} & \hc{0.935} & \hc{0.508} & \hc{0.498} & \hh{0.020} & 1.00 & \hcost{0.006} \\
  & Hybrid   & \hc{0.315} & \hc{0.848} & \hc{0.536} & \hc{0.951} & \hc{0.522} & \hc{0.537} & \hh{0.010} & 1.00 & \hcost{0.005} \\
  & Agentic  & \hcb{0.355} & \hcb{0.866} & \hcb{0.612} & \hc{0.638} & \hcb{0.537} & \hc{0.510} & \hh{0.390} & 3.02 & \hcost{0.045} \\
  & \corpusskill{}$^{\ddagger}$ & \hc{0.240} & \hc{0.847} & \hc{0.388} & \hc{0.706} & \hc{0.367} & \hc{0.499} & \hh{0.285} & 6.92 & \hcost{1.475} \\
\midrule
\multirow{5}{*}{CUAD}
  & BM25     & \hcb{0.314} & \hc{0.824} & \hc{0.277} & \hc{0.962} & \hc{0.181} & \hc{0.477} & \hhb{0.000} & 1.00 & \hcostb{0.007} \\
  & Dense    & \hc{0.304} & \hc{0.819} & \hcb{0.279} & \hcb{0.963} & \hc{0.167} & \hcb{0.742} & \hh{0.005} & 1.00 & \hcost{0.008} \\
  & Hybrid   & \hc{0.308} & \hc{0.821} & \hc{0.277} & \hc{0.945} & \hc{0.195} & \hc{0.685} & \hhb{0.000} & 1.00 & \hcost{0.007} \\
  & Agentic  & \hc{0.249} & \hc{0.819} & \hc{0.270} & \hc{0.402} & \hc{0.177} & \hc{0.706} & \hh{0.695} & 4.27 & \hcost{0.080} \\
  & \corpusskill{}$^{\ddagger}$ & \hc{0.272} & \hcb{0.828} & \hc{0.237} & \hc{0.643} & \hcb{0.217} & \hc{0.439} & \hh{0.430} & 5.80 & \hcost{0.502} \\
\bottomrule
\end{tabular}
\vspace{-0.5em}
\end{table*}

\begin{table}[!t]
\centering
\scriptsize
\setlength{\tabcolsep}{2.5pt}
\renewcommand{\arraystretch}{1.0}
\caption{Ablations on WixQA: \emph{(a)} cluster structure, \emph{(b)} agent exploration budget, \emph{(c)} serving LLM (all 200 queries), \emph{(d)} compiler LLM (50-query subset; compile cost \$9.5 vs.\ \$1.97). Per-panel best in \textbf{bold}.}
\label{tab:ablation}
\begin{tabular}{l cccccc r}
\toprule
\textbf{Variant} & F1$\uparrow$ & BERT$\uparrow$ & Fact.$\uparrow$ & CtxR$\uparrow$ & Halluc.$\downarrow$ & Turns & \$/q$\downarrow$ \\
\midrule
\multicolumn{8}{l}{\textit{(a) Cluster structure ($p$/skills/depth)}} \\
Narrow (5/2/5)   & \hcb{0.462} & \hcb{0.864} & \hc{0.758} & \hc{0.690} & \hh{0.080} & 2.38 & \hcostb{0.133} \\
Default (10/6/3) & \hc{0.456} & \hc{0.862} & \hcb{0.767} & \hcb{0.708} & \hhb{0.045} & 2.38 & \hcost{0.153} \\
Wide (20/4/3)    & \hc{0.382} & \hc{0.847} & \hc{0.552} & \hc{0.414} & \hh{0.245} & 3.69 & \hcost{0.565} \\
\midrule
\multicolumn{8}{l}{\textit{(b) Agent exploration budget (max rounds)}} \\
5 rounds  & \hc{0.455} & \hc{0.863} & \hcb{0.774} & \hc{0.702} & \hh{0.065} & 2.35 & \hcost{0.145} \\
10 rounds & \hcb{0.459} & \hcb{0.863} & \hc{0.754} & \hc{0.693} & \hh{0.065} & 2.36 & \hcostb{0.142} \\
20 rounds & \hc{0.456} & \hc{0.862} & \hc{0.767} & \hcb{0.708} & \hhb{0.045} & 2.38 & \hcost{0.153} \\
\midrule
\multicolumn{8}{l}{\textit{(c) Serving LLM (default tree, max\_turns${=}20$)}} \\
Sonnet 4.6 & \hcb{0.456} & \hcb{0.862} & \hcb{0.767} & \hc{0.708} & \hhb{0.045} & 2.38 & \hcost{0.153} \\
Haiku 4.5  & \hc{0.420} & \hc{0.856} & \hc{0.671} & \hcb{0.722} & \hh{0.180} & 2.92 & \hcostb{0.093} \\
\midrule
\multicolumn{8}{l}{\textit{(d) Compiler LLM (same serve setup, 50-query subset)}} \\
Sonnet 4.6 & \hcb{0.433} & \hcb{0.864} & -- & \hc{0.665} & \hh{0.082} & -- & -- \\
Haiku 4.5  & \hc{0.421} & \hcb{0.864} & -- & \hcb{0.688} & \hhb{0.020} & -- & -- \\
\bottomrule
\end{tabular}
\vspace{-0.5em}
\end{table}

\subsection{Setup}
\label{sec:setup}

\paragraph{Dataset and compilation.}
We evaluate on WixQA \citep{cohen2025wixqa}, a customer-support QA benchmark with 6{,}221 support articles and 200 expert-written test queries with gold answers and gold article IDs. We compile WixQA with branching ratio $p{=}10$ and maximum top-level clusters $K{=}10$, producing a 3-level hierarchy with 6 top-level skills and 665 navigation files. Compilation takes 19.5 minutes on an RTX 4090 server.

\paragraph{Baselines.}
All methods use the same answer LLM and the same evaluation protocol. We compare five baselines spanning three retrieval paradigms: \textbf{BM25} (sparse keyword retrieval \citep{robertson2009bm25}), \textbf{Dense} (Qwen3-Embedding-0.6B \citep{zhang2025qwen3embed} with a FAISS index \citep{johnson2019faiss}), \textbf{Hybrid} (RRF \citep{cormack2009rrf} of BM25 and Dense), \textbf{RAPTOR} (hierarchical tree retrieval \citep{sarthi2024raptor} with UMAP+GMM+BIC clustering), and \textbf{Agentic} (an LLM agent with iterative access to BM25, Dense, and Hybrid tools). The first three are single-shot (top-5 passages, one LLM call). RAPTOR uses collapsed tree retrieval with the \emph{same} summarization LLM as our compiler (Claude Sonnet~4.6), and Agentic is allowed up to 10 rounds. \Cref{sec:ablations} additionally controls for the compile-time encoder in both directions. Full implementation in \Cref{app:baselines}.

\paragraph{Metrics.}
We use a seven-metric suite. Quality: \emph{Token F1} (token precision/recall harmonic mean against gold), \emph{BERTScore-F1} \citep{zhang2020bertscore}, and \emph{Factuality} (LLM-as-judge \citep{zheng2023llmjudge} 1--5 vs.\ gold, normalized to 0--1). Grounding: three RAGAS-style \citep{es2024ragas} LLM-judge metrics (\emph{Faithfulness$_{\text{grd}}$}, \emph{Context Recall} (CtxR), \emph{Context Precision} (CtxP)) plus \emph{Hallucination Rate}, the fraction of queries with Faithfulness$_{\text{grd}}<0.6$. We also report turn count and \$/query. Cross-method fairness details are in \Cref{app:metrics}. Every win/tie/loss statement follows one rule: a dataset is a \emph{win} if \corpusskill{}'s Token F1 exceeds the best baseline's under a per-query paired bootstrap (10{,}000 resamples) at $p{<}0.05$, a \emph{loss} in reverse, and a \emph{tie} otherwise; per-dataset deltas and $p$-values are in \Cref{app:stats}.

\subsection{Main Results}
\label{sec:results}

\paragraph{Answer quality.}
\textbf{\corpusskill{} achieves the highest score on every answer-quality metric and on both retrieval-coverage metrics} (\Cref{tab:main_results}). Token F1 reaches 0.456 (a 21\% relative gain over Agentic 0.378 and 25\% over Dense 0.364). Three independent serve passes yield Token F1 $0.458 \pm 0.002$, run-to-run noise roughly $30\times$ smaller than the gap to the strongest baseline, and the lead over RAPTOR is significant ($p{<}0.0001$; \Cref{app:stats}). BERTScore-F1 is 0.862, ahead of Agentic (0.843), RAPTOR (0.839), and the flat retrievers (0.830--0.833). Factuality 0.767 exceeds the next-best methods Agentic (0.705) and RAPTOR (0.704) by 6 absolute points. CtxR/CtxP lead by clear margins (0.708/0.829 vs.\ 0.618/0.659 for RAPTOR and 0.452/0.649 for Agentic). Both hierarchical methods outperform the flat baselines, but \corpusskill{} extends RAPTOR's gains by a further 9\% on Factuality and 15\% on CtxR: cluster summaries surface thematic connections flat embeddings miss \citep{sun2026position}, letting the agent match queries to relevant document \emph{groups} rather than passages.

\paragraph{Faithfulness and hallucination.}
Single-shot retrievers score Faithfulness$_{\text{grd}}$ near 0.83--0.91 with Hallucination $\approx$0\% by construction (one passage in, little room to drift). Multi-turn agents face a strictly harder grounding setting because any claim not exactly supported by an accumulated document is flagged, and Agentic suffers severely (0.528 / 50\%). \corpusskill{} closes most of this gap (0.859 / 4.5\%, within 0.05 of the best single-shot baseline) while keeping the answer-quality lead.

\paragraph{Cost.}
With Anthropic ephemeral prompt caching on the navigation context, \corpusskill{} costs \textbf{\$0.153/query}, roughly 1.9$\times$ Agentic (\$0.082) and 13--22$\times$ the single-shot retrievers (\$0.007--\$0.012); removing the cache discount raises this to \$0.302 without changing the ordering (\Cref{app:stats}). The Skills API loads navigation files in every call, so cached input tokens dominate. Typical queries reuse $\sim$70\% of the prompt across consecutive turns and 100\% across queries that target the same top-level branch (billed at one-tenth the base rate once cached). The ablations (\Cref{sec:ablations}) show that Haiku-4.5 or a tighter exploration budget cut cost further while keeping most of the quality lead. Two qualitative navigation patterns recur in the traces: \emph{direct descent} (one \skillmd{}, one \indexmd{}, one document, stop) and \emph{cross-branch exploration} via \texttt{entity\_index.json} or the \texttt{\#\# Related skills} block (full traces in \Cref{app:traces}).

\subsection{Generalization to RAGBench}
\label{sec:ragbench}

To probe how \corpusskill{} behaves outside its target domain, we run the same five methods on ten RAGBench~\citep{friel2024ragbench} subsets (CovidQA~\citep{moller2020covidqa}, DelucionQA~\citep{sadat2023delucionqa}, Emanual~\citep{nandy2021emanual}, ExpertQA~\citep{malaviya2024expertqa}, TechQA~\citep{castelli2020techqa}, FinQA~\citep{chen2021finqa}, CUAD~\citep{hendrycks2021cuad}, HAGRID~\citep{kamalloo2023hagrid}, TatQA~\citep{zhu2021tatqa}, ClapNQ~\citep{rosenthal2025clapnq}), sampling 200 test queries per subset (or all available). RAGBench is published as a generator-evaluation benchmark with per-row gold contexts. To convert it into a retrieval benchmark we build the \emph{expanded retrieval pool} as the union of all train+val+test documents (deduplicated by content hash, pool sizes 221--12{,}727). This is intentionally harder than the original RAGBench protocol: every test query competes against thousands of distractors.

\paragraph{Corpus navigation is not one-size-fits-all.}
Under the paired-bootstrap criterion of \Cref{sec:setup}, \corpusskill{} wins on 5 of the 11 datasets (WixQA $+$0.059, Emanual $+$0.061, DelucionQA $+$0.034, ExpertQA $+$0.027, ClapNQ $+$0.135; all $p{\le}0.010$), ties on 3, and loses on 3 (\Cref{tab:ragbench_results}, per-dataset statistics in \Cref{app:stats}). The split is not random but instead tracks the \emph{structure} of the underlying corpus, and three regimes separate wins from losses. (i)~\emph{Single-domain, topically coherent corpora} where each document covers one feature or topic (WixQA, Emanual, DelucionQA, TechQA, ExpertQA, ClapNQ): \corpusskill{} leads every retrieval baseline, significantly so on all but TechQA ($+$0.013, $p{=}0.14$), with ClapNQ the most striking case (0.468 vs.\ Dense 0.333, $+$0.135 absolute). (ii)~\emph{Heterogeneous or numerical-table corpora} (CovidQA, FinQA): all methods sit within 0.01 F1 of the best baseline ($p{\ge}0.33$), and these are statistical ties. (iii)~\emph{Open-domain factoid or homogeneous-tabular corpora} (HAGRID, TatQA, CUAD): \corpusskill{} loses by 0.05--0.17 F1 ($p{<}0.001$) because the level-1 cluster summaries collapse onto generic labels (HAGRID's open-domain mix), describe near-identical units (TatQA's repeated SEC tables), or partition a single long document across multiple clause types (CUAD).

\paragraph{Grounding and positioning.}
On grounding, \corpusskill{}'s per-dataset Faithfulness$_{\text{grd}}$ / Hallucination Rate sit between single-shot retrievers (which feed back one grounded passage by construction) and the Agentic baseline, consistently halving Agentic's hallucination rate while staying competitive with single-shot grounding on the datasets where it leads on F1. We therefore position \corpusskill{} as a complementary primitive rather than a universal replacement for retrieval: preferred for curated, single-domain knowledge bases (this paper's deployment target), with flat retrieval or RAPTOR retained as the default for open-domain or homogeneous-table pools. The HAGRID/CUAD/TatQA losses are structural, and motivate the follow-on directions in the Limitations.

\subsection{Ablation Studies}
\label{sec:ablations}

\begin{table}[!t]
\centering
\scriptsize
\setlength{\tabcolsep}{3pt}
\renewcommand{\arraystretch}{1.0}
\caption{Compile-enrichment decomposition on WixQA (200 queries per variant); top group: compile-resource components, bottom: navigational structure.}
\label{tab:decomposition}
\begin{tabular}{l cccc}
\toprule
\textbf{Variant} & F1$\uparrow$ & Fact.$\uparrow$ & CtxR$\uparrow$ & CtxP$\uparrow$ \\
\midrule
Full & \hc{0.456} & \hc{0.767} & \hc{0.708} & \hc{0.829} \\
\midrule
$-$ verify-and-repartition & \hc{0.456} & \hc{0.760} & \hc{0.677} & \hc{0.821} \\
$-$ doc summary cards      & \hc{0.463} & \hc{0.780} & \hc{0.695} & \hc{0.811} \\
$-$ 8B encoder (use 0.6B)  & \hc{0.461} & \hc{0.751} & \hc{0.702} & \hc{0.840} \\
\midrule
$-$ exemplars              & \hc{0.401} & \hc{0.609} & \hc{0.514} & \hc{0.676} \\
$-$ soft assignment        & \hc{0.333} & \hc{0.397} & \hc{0.336} & \hc{0.413} \\
$-$ entity index           & \hc{0.312} & \hc{0.359} & \hc{0.278} & \hc{0.344} \\
\bottomrule
\end{tabular}
\vspace{-0.5em}
\end{table}

We ablate compile-time and serve-time design choices on WixQA: cluster structure, exploration budget, serving and compiler LLMs (\Cref{tab:ablation}), encoder choice, and the compile-time enrichments (\Cref{tab:decomposition}). All variants share the two-tool agent configuration and are directly comparable to \Cref{tab:main_results}.

\paragraph{Cluster structure (panel~a).}
Varying $p$ at $K{=}10$ produces narrow ($p{=}5$), default ($p{=}10$), and wide ($p{=}20$) trees. Narrow and default land within $\approx$1 F1 point (0.462 vs.\ 0.456) on every quality metric, and narrow is slightly cheaper (\$0.133 vs.\ \$0.153). The \emph{wide} shape collapses to F1 0.382 ($-0.07$), CtxR 0.414 ($-0.29$), Hallucination 24.5\%, 3.7 turns / \$0.57 per query: \textbf{once a level-1 cluster must summarize hundreds of topically-mixed documents its summary becomes too generic to route on}, reproducing the TatQA failure mode of \Cref{sec:ragbench} in miniature.

\paragraph{Agent exploration budget (panel~b).}
Varying max rounds (5/10/20) has minimal impact on quality (within 0.005 F1, avg.\ 2.4 turns regardless of budget) because the hierarchy is well-organized enough that the agent rarely needs extended exploration. A larger budget mainly helps grounding: Hallucination drops monotonically from 6.5\% (budget 5) to 4.5\% (budget 20), suggesting that the extra rounds are used to verify rather than re-search.

\paragraph{Serving LLM choice (panel~c).}
Swapping Sonnet~4.6 for Haiku~4.5 (same compiled tree) cuts cost \$0.153 $\to$ \$0.093 ($-39\%$) while retaining 92\% of Sonnet's F1 (0.420) and 87\% of its Factuality (0.671). CtxR actually \emph{increases} (0.708 $\to$ 0.722) because Haiku tends to read one more document before answering. The tradeoff is grounding: Haiku's Hallucination Rate is $4\times$ Sonnet's, so Sonnet remains the right choice for compliance-sensitive deployments. \textbf{The Haiku variant is the cheapest \corpusskill{} configuration we measure and still substantially beats every flat retrieval baseline on F1 by 2--6 absolute points}, evidence that the compiled tree, not the navigator's sophistication, drives retrieval performance.

\paragraph{Compiler LLM (panel~d).}
Swapping the \emph{entire} compile LLM from Sonnet~4.6 to Haiku~4.5 retains 97\% of F1 at one-fifth the compile cost: neither side of the pipeline needs a frontier model. Compile-time summarization errors are also contained: summaries only \emph{route} the agent, while answers are grounded in the retrieved raw documents, so a weak summary can cost recall but cannot inject unsupported content (cf.\ \Cref{app:failure_analysis}).

\paragraph{Encoder choices.}
\corpusskill{} compiles with Qwen3-Embedding-8B; the baselines index with the 0.6B variant (\Cref{app:baselines}). A $2{\times}2$ encoder cross rules this asymmetry out: compiled with the baselines' 0.6B encoder, \corpusskill{} scores 0.461 F1 vs.\ $0.458 \pm 0.002$ for the 8B compile ($p{=}0.37$), while the $13\times$ larger encoder helps the baselines by at most $+$0.023 F1. Against RAPTOR, whose recursive summarization already uses the same Sonnet~4.6 summarizer as our compiler, \corpusskill{} leads by $+$0.064 F1 matched at 0.6B and $+$0.035 at 8B (both $p{<}0.0001$; full tables in \Cref{app:stats}).

\paragraph{Component decomposition.}
Removing one compile-time enrichment at a time (\Cref{tab:decomposition}) separates the two groups cleanly: dropping the entity index takes F1 from 0.456 to 0.312, soft assignment to 0.333, and exemplars to 0.401, while verify-and-repartition is F1-neutral and removing the summary cards or the 8B encoder shifts F1 by at most 0.007 (n.s.). The gains come from the navigable structure; compile-side resources contribute little.

\subsection{Scaling to 50k Documents}
\label{sec:scale}

\begin{table}[!t]
\centering
\scriptsize
\setlength{\tabcolsep}{4pt}
\renewcommand{\arraystretch}{1.0}
\caption{Token F1 on CovidQA as its pool grows 5k $\to$ 50{,}000 documents (same 200 queries and compiler). The 50k compile costs \$54.1 / 76\,min (\$1.08/1k docs vs.\ \$3.33 at 6k) and adds one hierarchy level.}
\label{tab:scale}
\begin{tabular}{l ccc}
\toprule
\textbf{Method} & 5k pool & 50k pool & $\Delta$ \\
\midrule
\corpusskill{} & \hc{0.208} & \hc{0.191} & $-$0.017 \\
Dense          & \hc{0.215} & \hc{0.187} & $-$0.028 \\
Hybrid         & \hc{0.216} & \hc{0.175} & $-$0.041 \\
BM25           & \hc{0.204} & \hc{0.169} & $-$0.035 \\
\bottomrule
\end{tabular}
\vspace{-0.5em}
\end{table}

Growing CovidQA's pool to 50{,}000 documents with in-domain CORD-19 abstracts ($8\times$ our largest corpus, same compiler and protocol) adds exactly one hierarchy level, matching the $O(\log_p N)$ prediction, at sub-linear compile cost (\Cref{tab:scale}); all 200 queries navigate the larger tree without error, and the compact-mode output fits production Skills API limits. With 45{,}000 same-domain distractors in the pool, every method degrades; navigation degrades least, leaving \corpusskill{} highest in F1, tied with Dense ($p{=}0.74$) and above BM25 ($p{=}0.04$), so CovidQA remains in the tie regime at $10\times$ scale.

\FloatBarrier
\section{Conclusion}
\label{sec:conclusion}

We introduced \corpusskill{}, a system-level retrieval architecture that compiles a document corpus into a hierarchical skill directory and replaces embedding-based retrieval with agent-driven navigation. On WixQA, it improves answer quality and grounding over flat, hierarchical, and agentic baselines. An eleven-dataset study bounds the scope: navigation significantly wins on five datasets, ties on three, and loses on three. It suits single-domain collections with a recoverable taxonomy, while flat retrieval remains preferable where top-level clustering collapses. Controlled ablations trace the gains to the compiled tree and its enrichments, and these conclusions hold at 50{,}000 documents. Because the output is plain skill directories requiring no vector database, it integrates directly into existing agent deployments. For bounded corpora with recoverable topical structure, making a corpus \emph{navigable} complements making it \emph{searchable}.

\section*{Limitations}
\label{sec:limitations}
\addcontentsline{toc}{section}{Limitations}

\paragraph{Cost.}
Navigation file content is included in every API call, so input tokens accumulate across turns. With Anthropic ephemeral prompt caching (used throughout our results), \corpusskill{} costs \$0.153/query, about 1.9$\times$ the Agentic baseline; without caching this roughly doubles to \$0.302/query, though the ordering against the baselines is unchanged (\Cref{app:stats}). Swapping to Haiku~4.5 reduces the cached cost to \$0.093 while retaining most of the quality lead (\Cref{sec:ablations}). The method therefore targets high-value queries where improved grounding justifies the cost premium. For high-volume, low-stakes queries, single-shot retrieval remains more cost-effective.

\paragraph{Routing errors.}
Although soft assignment and entity cross-linking (\Cref{sec:compile}) let the agent discover cross-cutting content, our failure analysis (\Cref{app:failure_analysis}) shows that top-level routing errors still account for the majority of failures. Documents spanning multiple topics can be cross-referenced in secondary branches, but the primary routing decision remains a bottleneck when the query targets the less obvious topic.

\paragraph{Compilation and corpus churn.}
The one-time compilation takes $\sim$20 minutes of wall-clock time for WixQA (6{,}221 documents), negligible for enterprise deployments handling thousands of daily queries, and per-document cost falls sub-linearly with corpus size (\Cref{sec:scale}). The system assumes a mostly static corpus: incremental insertion, deletion, and hierarchy repair are not supported, so any change requires recompilation, and frequently changing collections face recurring compile costs. The cheaper Haiku~4.5 compiler (97\% of F1 at one-fifth the cost, \Cref{sec:ablations}) lowers the cost of each recompilation.

\paragraph{Corpus scope.}
Our RAGBench study (\Cref{sec:ragbench}) identifies three corpus shapes where navigation does not help: long extractive documents where answers are verbatim spans (CUAD), homogeneous tabular corpora whose cluster summaries collapse onto near-duplicate labels (TatQA), and open-domain pools where dense retrieval already surfaces sufficient evidence in a single shot (HAGRID). These are structural scope limits rather than implementation issues, and motivate hybrid per-query routing between navigation and retrieval.

\bibliography{references}

\appendix
\crefalias{section}{appendix}
\crefalias{subsection}{appendix}
\section{Compilation Details}
\label{app:compilation}

\paragraph{Top-level skill labels.}
The 6 skills produced by the default configuration ($p{=}10$, $K{=}10$) on the WixQA corpus are shown in \Cref{tab:skill_labels}.

\begin{table}[h]
\centering
\small
\setlength{\tabcolsep}{5pt}
\renewcommand{\arraystretch}{1.15}
\caption{Top-level skills produced by \corpusskill{} on WixQA.}
\label{tab:skill_labels}
\begin{tabular}{@{}lrl@{}}
\toprule
\textbf{Skill Label} & \textbf{Docs} & \textbf{Coverage} \\
\midrule
commerce-operations & 1{,}682 & Commerce, POS, payments \\
platform-overview & 1{,}770 & Editors, CMS, design \\
platform-ecosystem & 1{,}797 & Domains, billing, lifecycle \\
members-community & 256 & Members, gated access \\
bookings-platform & 283 & Bookings, scheduling \\
media-display & 433 & Galleries, storefront \\
\bottomrule
\end{tabular}
\vspace{-0.5em}
\end{table}

\paragraph{Embedding.}
Documents and cluster summaries are embedded using Qwen/Qwen3-Embedding-8B, a 7.6B-parameter sentence encoder producing 4{,}096-dimensional embeddings.
We use the document prompt template with batch size 32 on a single GPU.

\paragraph{Document summary cards.}
Before embedding, each document is processed by a lightweight \emph{summary-card} LLM call (Claude Haiku~4.5, $\le$250 output tokens, single API call per document, async with a semaphore of 20).
The card is a JSON object with fields \texttt{title}, \texttt{one\_line}, and \texttt{phrases} (up to four 1--4-word distinctive phrases such as product names or feature flags).
The card is concatenated with the (truncated) raw text as a single string and embedded as one vector per document; the card is also retained on the document record and reused later as INDEX.md row metadata, so the per-document summarization cost is amortized across embedding, navigation, and serving.

\paragraph{Cluster summarization.}
Cluster summaries are produced by Claude Sonnet~4.6 with direct API calls (no agent loop).
Level-1 summaries condense up to 15 document texts (600 characters each); deeper-level summaries condense up to 20 child summaries (300 characters each).
Summarization is performed concurrently using async calls with a semaphore of 20.
Internal-level clusters are re-embedded by concatenating their summary with a sampled set of member excerpts before passing the result to the next round of K-Means, so each higher-level vector reflects both the abstractive summary and surface-lexical signal.

\paragraph{Soft assignment.}
K-Means assignment uses a margin threshold $\tau{=}0.05$ on cosine similarity: items whose top-1 versus top-2 centroid similarity gap is below $\tau$ retain their primary parent and additionally register a secondary parent.
Secondary assignments are not duplicated as full documents; the runner-up parent's \indexmd{} lists them as \texttt{@see} stubs pointing back to the canonical location.
On the WixQA compile this produces 2{,}068 secondary stubs spread across 665 navigation files.

\paragraph{LLM verify-and-repartition.}
After each level's clusters are summarized, a verify-and-repartition LLM pass examines all sibling cluster summaries together (in batches of up to 60 clusters per call) and proposes ID-level reassignments where two summaries cover the same topic (\emph{confusable}) or one summary spans multiple topics (\emph{mixed}).
The pass operates at levels 1 and 2 only and visits each cluster at most once per compile; affected clusters are re-summarized after reassignment.
This adds one extra Sonnet call per affected sibling group and a re-summarization call per repartitioned cluster.

\paragraph{Exemplars.}
For every non-leaf cluster, the top-3 documents closest (by cosine similarity to the cluster centroid) are surfaced under an \texttt{\#\# Example documents} section in the rendered \skillmd{}/\indexmd{}.
Exemplars carry the document title and one-line description from the summary card.

\paragraph{Entity extraction and related skills.}
Once the cluster hierarchy is stable, a single batched Sonnet call set tags each cluster summary with up to 12 named entities (proper nouns, product names, error codes, file formats) and up to 4 document types.
The entity tags are aggregated into a corpus-level \texttt{entity\_index.json} that maps each entity to the skill paths that mention it (sorted by mention count).
For every top-level skill, the top-3 sibling skills by entity-Jaccard overlap are recorded as \texttt{\#\# Related skills}, giving the agent explicit cross-jump targets.
On the WixQA compile this produces 2{,}328 distinct entries in \texttt{entity\_index.json}.

\paragraph{File counts.}
The compilation produces 665 navigation files (\skillmd{} + \indexmd{}), a \texttt{documents.json} file storing the full text of all 6,221 documents (13~MB), and a \texttt{entity\_index.json} mapping 2{,}328 entities to skill paths. Total output size is 16~MB.

\paragraph{Navigation complexity.}
\label{app:complexity}
The hierarchy depth grows as $O(\log_p N)$: with branching ratio $p$ and $N$ documents, the agent traverses at most $L = \lceil \log_p N \rceil$ levels to reach a leaf. At each level $\ell$, the agent reads one navigation file summarizing $O(p)$ children. A single decision at level $\ell$ reduces the candidate set from $N/p^{\ell-1}$ to $N/p^{\ell}$, yielding a multiplicative information funnel:
\begin{equation}
\label{eq:funnel}
N \;\xrightarrow{\div\, p}\; \frac{N}{p} \;\xrightarrow{\div\, p}\; \frac{N}{p^2} \;\cdots\; \xrightarrow{\div\, p}\; \frac{N}{p^L} \approx 1
\end{equation}
The total summaries inspected scale as $O(p \cdot \log_p N)$ rather than $O(N)$. For WixQA ($N{=}6{,}221$, $p{=}10$), this gives $L{=}3$ levels and approximately 30 summaries to locate a single document among thousands. Note that this bound counts summaries inspected under an idealized single-path traversal; in practice, the actual token cost is higher because the Skills API includes previously loaded navigation files in each subsequent call, and the agent may explore multiple branches. The per-query costs reported in \Cref{sec:experiments} reflect these real-world overheads. A 100{,}000-document corpus with $p{=}10$ would require only one additional level ($L{=}5$).

\paragraph{Document-level summarization prompt.}
Used at the lowest level to summarize a cluster of raw documents.

\begin{scriptsize}
\begin{verbatim}
You are summarizing a cluster of related
documents from a knowledge base. Write a
2-3 sentence summary that captures:
1. The common TOPIC area these documents
   cover
2. The types of QUESTIONS these documents
   answer
3. Key TERMS or features mentioned across
   documents

Be specific and concrete. Name actual
features, products, or processes.

Documents ({N} total, showing up to 15):

--- Document 1 ---
{doc_text[:600]}

--- Document 2 ---
{doc_text[:600]}
[...]

Summary:
\end{verbatim}
\end{scriptsize}

\paragraph{Cluster-level summarization prompt.}
Used at higher levels to summarize a group of sub-cluster summaries into a broader overview.

\begin{scriptsize}
\begin{verbatim}
You are summarizing a level-{L} grouping
of {N} sub-groups from a knowledge base.
Each sub-group already has a summary below.

Write a 2-3 sentence overview that captures:
1. The broad DOMAIN these sub-groups cover
2. The range of TOPICS within this domain
3. What types of user QUESTIONS this group
   can answer

Be specific -- name the main product areas,
features, or workflows.

Sub-group summaries:

- Sub-group 1: {summary[:300]}
- Sub-group 2: {summary[:300]}
[...]

Overview:
\end{verbatim}
\end{scriptsize}

\paragraph{Labeling prompt.}
Used to generate a filesystem-safe directory name from a cluster summary. The response is post-processed: lowercased, non-alphanumeric characters replaced with hyphens, truncated to 50 characters.

\begin{scriptsize}
\begin{verbatim}
Generate a short (2-5 word) filesystem-safe
label for this cluster. Use lowercase,
hyphens instead of spaces. No quotes.

Summary: {summary[:500]}
\end{verbatim}
\end{scriptsize}

\section{Skill Structure Examples}
\label{app:skill_structure}

We show the full content of representative navigation files from the WixQA compilation (default configuration: $p{=}10$, $K{=}10$).
These files are what the agent reads during navigation.

\paragraph{Root-level \skillmd{}.}
This file sits at the top of the \texttt{skill-00-wix-commerce-operations} skill directory, the largest skill covering 1,682 documents across 11 subgroups.
The agent reads this file in a single tool call and uses the subgroup summaries to decide which branch to explore.

\begin{scriptsize}
\begin{verbatim}
---
name: skill-00-wix-commerce-operations
description: >
  This group covers the Wix commerce and
  business operations ecosystem, spanning
  the full range of tools for running
  transactions, managing services, and
  scaling a business on Wix. Topics include
  e-commerce store management, restaurant
  and hotel operations, event ticketing,
  point-of-sale systems, bookings and
  scheduling, pricing plans, payments and
  payment providers, tax and invoicing,
  promotions and discounts...
level: 3
num_documents: 1682
---

## Overview

This group covers the Wix commerce and
business operations ecosystem, spanning
the full range of tools for running
transactions, managing services, and
scaling a business on the Wix platform.
Topics include e-commerce, restaurant and
hotel operations, event ticketing, POS,
bookings, pricing plans, payments, tax,
invoicing, promotions, and multilingual
localization. Users can get answers about
setting up and configuring any Wix commerce
product and processing payments across
global providers.

## Contents

### Sub-groups (directories)

Read the INDEX.md in each sub-group to
understand what it covers.

- **group-00-wix-events-platform/**
  (138 docs): Wix Events platform: event
  creation, ticketing, registration, guest
  management, and promotion.
- **group-01-wix-restaurants-platform/**
  (232 docs): Wix Restaurants: menus,
  ordering, reservations, delivery, POS
  integrations, and fulfillment.
- **group-02-wix-hotels-bookings/**
  (109 docs): Wix Hotels and Bookings:
  room configuration, pricing, calendar
  sync, and OTA channel management.
- **group-03-wix-pos-ecosystem/**
  (167 docs): Wix POS ecosystem: hardware
  setup, payment processing, staff
  management, and inventory.
- **group-04-wix-payments-processing/**
  (113 docs): Payments lifecycle: account
  setup, chargebacks, fraud prevention,
  payouts, and subscription billing.
- **group-05-wix-billing-tax-management/**
  (110 docs): Financial and tax management:
  invoicing, quotes, Avalara, VAT/GST.
[... 5 more sub-groups omitted ...]
\end{verbatim}
\end{scriptsize}

\paragraph{Mid-level \indexmd{}.}
This file sits one level below the root, inside \texttt{group-04-wix-payments-processing/}, covering 113 documents across 10 leaf groups.
The agent reads this after deciding from the \skillmd{} above that payments are relevant.

\begin{scriptsize}
\begin{verbatim}
---
name: wix-payments-processing
description: >
  Covers the broad domain of payments on
  the Wix platform, spanning the full
  lifecycle from account setup to
  transaction management, payouts, and
  compliance.
level: 2
num_documents: 113
---

## Overview

Topics range from setting up and verifying
Wix Payments accounts, enabling third-party
providers (PayPal, Stripe, Apple Pay,
Google Pay), and managing chargebacks and
fraud prevention, to payout schedules,
prohibited product restrictions,
subscription billing, and unsupported
payment features.

## Contents

### Sub-groups (directories)

- **group-00-wix-payments-chargebacks/**
  (12 docs): Chargeback management,
  transaction fees, and merchant practices.
- **group-01-wix-payment-setup/**
  (22 docs): Setup and troubleshooting of
  payment methods across commerce products.
- **group-02-wix-payments-payouts/**
  (15 docs): Payout management, settlement
  reports, and fund transfer timelines.
- **group-03-payment-flexibility-requests/**
  (9 docs): Feature requests for split,
  installment, and deferred payments.
- **group-04-wix-paypal-button-setup/**
  (7 docs): PayPal Buy Now and Donate
  button configuration.
- **group-05-wix-payments-account-setup/**
  (13 docs): Account setup, verification,
  and EIN/identity documents.
- **group-06-wix-payments-processing/**
  (10 docs): Payment processing issues,
  fraud prevention, and declined payments.
- **group-07-wix-payments-management/**
  (9 docs): Transaction dashboards, payment
  history, and settlement reports.
- **group-08-prohibited-payment-products/**
  (6 docs): Payment provider compliance and
  restricted product categories.
- **group-09-wix-subscription-billing/**
  (10 docs): Premium plan billing, renewal
  payments, and payment method updates.
\end{verbatim}
\end{scriptsize}

\paragraph{Leaf-level \indexmd{}.}
This file sits at the bottom of the hierarchy, inside \texttt{group-06-wix-payments-processing/} under the payments processing branch.
It lists individual document IDs with titles and summaries. The agent reads this to select which documents to retrieve via \texttt{get\_document}.

\begin{scriptsize}
\begin{verbatim}
---
name: wix-payments-processing
description: >
  Payment processing issues, fraud
  prevention, and transaction management
  within Wix Payments.
level: 1
num_documents: 10
---

## Overview

These documents cover payment processing
issues, fraud prevention, and transaction
management within Wix Payments. They answer
questions about why payments are declined
(AVS mismatches, CVV errors, 3-D Secure
failures, suspected fraud), how to resolve
double charges, and how EMV chip liability
shift rules affect merchants.

## Contents

### Documents (10 items)

Use get_document with the doc_id to
read full content.

- `56ae7642b2a8c2ea` -- Best Practices
  for Accepting Card Payments
- `88f6b5848f6f662d` -- Troubleshooting
  Payment Failures and Declined
  Transactions
- `d22b8cf6e599e0b8` -- Double Charge on
  Bank Statement Explanation
- `448d7947c5a22489` -- Wix Payments:
  Liability Shift and Chip Card
  Requirements
- `bb73d9cb90385f12` -- Wix Payments
  Declined Payment Suspected Fraud
[... 5 more documents omitted ...]
\end{verbatim}
\end{scriptsize}

\section{Compilation Prompts}
\label{app:compile_prompts}

The compilation pipeline uses three LLM prompts, each issued as a single-turn API call (no agent loop or tool use).

\paragraph{Document-level summarization.}
Used at the lowest level to summarize a cluster of raw documents.

\begin{scriptsize}
\begin{verbatim}
You are summarizing a cluster of related
documents from a knowledge base. Write a
2-3 sentence summary that captures:
1. The common TOPIC area these documents
   cover
2. The types of QUESTIONS these documents
   answer
3. Key TERMS or features mentioned across
   documents

Be specific and concrete. Name actual
features, products, or processes.

Documents ({N} total, showing up to 15):

--- Document 1 ---
{doc_text[:600]}

--- Document 2 ---
{doc_text[:600]}
[...]

Summary:
\end{verbatim}
\end{scriptsize}

\paragraph{Cluster-level summarization.}
Used at higher levels to summarize a group of sub-cluster summaries into a broader overview.

\begin{scriptsize}
\begin{verbatim}
You are summarizing a level-{L} grouping
of {N} sub-groups from a knowledge base.
Each sub-group already has a summary below.

Write a 2-3 sentence overview that captures:
1. The broad DOMAIN these sub-groups cover
2. The range of TOPICS within this domain
3. What types of user QUESTIONS this group
   can answer

Be specific -- name the main product areas,
features, or workflows.

Sub-group summaries:

- Sub-group 1: {summary[:300]}
- Sub-group 2: {summary[:300]}
[...]

Overview:
\end{verbatim}
\end{scriptsize}

\paragraph{Labeling.}
Used to generate a filesystem-safe directory name from a cluster summary. The response is post-processed: lowercased, non-alphanumeric characters replaced with hyphens, truncated to 50 characters.

\begin{scriptsize}
\begin{verbatim}
Generate a short (2-5 word) filesystem-safe
label for this cluster. Use lowercase,
hyphens instead of spaces. No quotes.

Summary: {summary[:500]}
\end{verbatim}
\end{scriptsize}

\paragraph{Document summary card.}
Issued once per document at the start of compilation. The returned JSON card is concatenated with the truncated raw text for embedding and is reused as INDEX.md row metadata.

\begin{scriptsize}
\begin{verbatim}
You are summarizing a single document from
a knowledge base. Produce a compact JSON
'card' describing the document so an agent
can scan many cards quickly and pick the
relevant one.

Return ONLY a JSON object with exactly
these keys:
  title:    short descriptive title
            (5-12 words). Reuse the
            document's own title if it has
            one, otherwise infer.
  one_line: a single sentence (max 25
            words) describing what the
            document covers and what kind
            of question it answers.
  phrases:  list of 2-4 short distinctive
            phrases (2-6 words each) that
            are characteristic of this
            document and unlikely to appear
            in unrelated documents (e.g.
            product names, schema names,
            specific entities, code
            identifiers, section headers).

Be specific and concrete. Do not include
generic phrases like 'knowledge base
article' or 'support document'.

Document:
{document}

JSON card:
\end{verbatim}
\end{scriptsize}

\paragraph{Verify-and-repartition.}
Issued once per hierarchy level (skipped beyond 60 sibling clusters per call). The model is asked to flag confusable cluster pairs and mixed individual clusters, and to return ID-level reassignments. Changes are applied verbatim; affected clusters are re-summarized.

\begin{scriptsize}
\begin{verbatim}
You are auditing a clustering of items at
level {level} of a navigable knowledge
hierarchy. Each cluster has a label, a
short summary, and a list of member item
IDs with brief titles.

Two failure modes to look for:
  (a) CONFUSABLE: two or more clusters
      whose labels/summaries describe
      essentially the same topic.
  (b) MIXED: a single cluster whose
      members visibly belong to multiple
      distinct sub-topics that should be
      split.

For each problematic GROUP (a set of
confusable cluster IDs, or a single mixed
cluster ID), output a fresh partition of
its members by ID. Use ONLY the IDs already
listed in that group; do not invent IDs or
move items across non-flagged boundaries.

Return ONLY a JSON object with this shape:
{ "changes": [
  { "old_cluster_ids": ["L1-C2", "L1-C7"],
    "new_partition": {
      "label_a": ["id1","id2"],
      "label_b": ["id3","id4"]
    }
  }
] }
If no changes are needed, return
{ "changes": [] }.

Clusters at level {level}:
{clusters_block}

JSON:
\end{verbatim}
\end{scriptsize}

\paragraph{Entity and doc-type extraction.}
Issued once per cluster, batched concurrently, after the hierarchy stabilizes. The returned named entities are aggregated into \texttt{entity\_index.json} and drive the per-skill \texttt{Related skills} cross-references.

\begin{scriptsize}
\begin{verbatim}
You are tagging a topic cluster from a
knowledge base with the entities and
document types it contains.

Extract ONLY two categories:
  named_entities: proper-noun product
    names, feature names, person names,
    place names, organization names,
    schema names, specific identifiers.
    Exclude generic concepts (e.g.
    'authentication', 'payments' are
    generic; 'Stripe Connect',
    'Wix Bookings' are named entities).
  doc_types: kinds of documents present
    in the cluster, e.g. 'FAQ', 'release
    notes', 'balance sheet', 'contract
    clause', 'API reference', 'tutorial',
    'changelog'.

Return ONLY a JSON object:
{ "named_entities": ["..."],
  "doc_types": ["..."] }
Aim for 3-10 named entities and 1-3 doc
types. Use lowercase, no duplicates.

Cluster label: {label}
Cluster summary: {summary}
Example item titles:
{titles_block}

JSON:
\end{verbatim}
\end{scriptsize}

\section{Serving System Prompt}
\label{app:system_prompt}

The serving agent receives the following system prompt:

\begin{scriptsize}
\begin{verbatim}
# Corpus-Grounded Support Agent
#   (Hierarchical Navigation)

You are a knowledge agent that answers
questions by navigating a hierarchical
skill directory. You explore a structured
file tree where documents are organized
into topic clusters, and you have access
to an entity cross-index and a leaf-level
search tool to triangulate evidence.

## Hard Rules

- Every factual claim must trace to a
  document you retrieved via get_document.
- SKILL.md / INDEX.md files are NAVIGATION
  AIDS -- they tell you where to look,
  not what to say.
- Never fabricate steps, URLs, prices, or
  specifics not found in documents.
- Never guess. If you cannot find relevant
  content after thorough exploration,
  say so.

## Skill directory layout

skill-XX-topic/
  SKILL.md          <- top-level summary,
                       exemplar docs,
                       related skills,
                       index
  group-YY-subtopic/
    INDEX.md        <- sub-group summary
                       + child index
    group-ZZ/
      INDEX.md      <- leaf summary +
                       document rows
                       (id + title +
                       phrases)

Each SKILL.md / INDEX.md may contain:
- ## Overview      -- what this skill
                       covers
- ## Related skills -- sibling skills that
                       share entities;
                       cross-jump here when
                       the current branch
                       is thin
- ## Entities & document types
                    -- quick scan of named
                       entities present
- ## Example documents in this skill
                    -- representative items
                       (read one of these
                       to learn what kind
                       of content lives
                       here)
- ## Contents       -- sub-groups and/or
                       document rows
- ### See also      -- @see stubs
                       (documents whose
                       primary entry is
                       elsewhere)

A corpus-level entity_index.json is also
present at the corpus root. It maps each
named entity to the skill paths that
mention it; use it to find ALL skills
relevant to a query entity rather than
only the first match.

## Navigation Strategy (multi-candidate)

1. SCAN at least 2 candidate top-level
   skills before committing to a path.
   - If the query mentions a specific
     entity, first
     `cat entity_index.json | grep -i
       <entity>`
     to find every skill that references
     it, and read the SKILL.md of the top
     2 by mention count.
   - Otherwise read the SKILL.md of the 2
     most-plausible top-level skills.
2. For each candidate, descend ONE level
   (read its first relevant INDEX.md)
   before pruning. Write down which
   candidates you keep and why.
3. At the leaf, use `cat ... | grep ...`
   against the INDEX.md doc rows to pick
   the most promising doc_ids -- the rows
   include the document title, one-liner,
   and key phrases, which is usually
   enough to triangulate without opening
   anything.
4. If two candidate paths return
   comparable evidence, retrieve documents
   from BOTH and explicitly compare. Pick
   the best answer; do NOT just take the
   first match.
5. If a candidate skill has a `## Related
   skills` block and your initial path is
   thin, cross-jump to a related skill
   before giving up.
6. Read at least one full document via
   get_document before answering.

## Tools

- Code execution: Use `ls`, `cat`,
  `cat ... | grep ...` to navigate the
  skills hierarchy and the
  entity_index.json.
- get_document(doc_id): Retrieve the full
  text of a document by its ID.

## Answer Format

- First sentence = direct answer.
  No preamble.
- Factual questions: 1-3 sentences
  (~80 words max).
- Procedural questions: numbered steps
  only (~150 words max).
- Plain text. No bold, headers, or
  dividers.
- One approach only. Do not present
  alternatives.
- Never add "contact support" or closing
  remarks.
\end{verbatim}
\end{scriptsize}

\section{Full Navigation Traces}
\label{app:traces}

We present complete navigation traces with raw tool calls for four queries from the WixQA evaluation set (200 expert-written test queries, Claude Sonnet~4.6 serving agent).

\paragraph{Trace 1: Direct descent (Factuality 1.0, CtxR 1.0).}
Query: \emph{``I need to switch my business type from sole prop to LLC in order to use an EIN.''}

\begin{scriptsize}
\begin{verbatim}
[Context: agent sees 6 skill descriptions]

--- Turn 1 ---
Agent thought: The query is about changing
business type for Wix Payments. The skill
"wix-commerce-operations" covers payments,
taxes, and business operations.

Tool: code_execution
  cat skill-00-wix-commerce-operations/
    SKILL.md

Result: 11 subgroups listed, including:
  - group-04-wix-payments-processing/
    (113 docs): "payments lifecycle: account
    setup, chargebacks, fraud prevention,
    payouts, and subscription billing"

Tool: code_execution
  cat group-04-wix-payments-processing/
    INDEX.md

Result: 10 leaf groups. Agent identifies
  group-05-wix-payments-account-setup
  covering account setup and verification.

Tool: code_execution
  cat group-05-wix-payments-account-setup/
    INDEX.md

Result: 13 documents listed, including:
  - d56cc797742e: "Switching Between
    Individual and Business Account Types
    in Wix Payments"

--- Turn 2 ---
Tool: get_document
  {"doc_id": "d56cc797742e7d0f"}

Result: Full article text (1,247 chars)
  describing that account type cannot be
  changed directly; must contact Wix
  Customer Care.

--- Answer ---
"You cannot change your account type
directly within Wix Payments. To switch
from sole proprietorship to LLC, contact
Wix Customer Care with your EIN and LLC
formation documents ready."

Navigation: 6221 -> 1682 -> 113 -> 13 -> 1
\end{verbatim}
\end{scriptsize}

\paragraph{Trace 2: Multi-document retrieval (Factuality 1.0, CtxR 1.0).}
Query: \emph{``How do I sync the hotel app with my calendars iCal link to allow visitors to book available dates?''}

\begin{scriptsize}
\begin{verbatim}
[Context: agent sees 6 skill descriptions]

--- Turn 1 ---
Agent thought: Hotels and calendar sync
fall under "wix-commerce-operations" which
covers hotel operations and bookings.

Tool: code_execution
  cat skill-00-wix-commerce-operations/
    SKILL.md

Result: 11 subgroups listed, including:
  - group-02-wix-hotels-bookings/
    (109 docs): "Wix Hotels and Bookings:
    room configuration, pricing, calendar
    sync, and OTA channel management"

Tool: code_execution
  cat group-02-wix-hotels-bookings/
    INDEX.md

Result: 8 leaf groups. Agent identifies
  group-07-hotel-channel-manager-setup
  covering external calendar integration.

Tool: code_execution
  cat group-07-hotel-channel-manager-setup/
    INDEX.md

Result: 17 documents listed, including:
  - 5f02fc2f8937: "Importing Reservations
    from External Calendar (iCal)"
  - 741830b917b4: "Connecting Airbnb iCal
    to Your Site"

--- Turn 2 ---
Tool: get_document
  {"doc_id": "5f02fc2f8937d16a"}

Result: Step-by-step iCal import guide
  for original Wix Hotels.

Tool: get_document
  {"doc_id": "741830b917b47d07"}

Result: Airbnb iCal connection guide for
  Wix Hotels by HotelRunner.

--- Answer ---
"For original Wix Hotels: Go to Hotels
tab > Reservations > iCal, select room
type, paste iCal URL, click Save. For
HotelRunner: go to dashboard, navigate
to channel connections, add Airbnb iCal
link for each property..."

Navigation: 6221 -> 1682 -> 109 -> 17 -> 2
\end{verbatim}
\end{scriptsize}

\paragraph{Trace 3: Mobile quick action (Factuality 1.0, CtxR 1.0).}
Query: \emph{``I need to temporarily remove the quick action button from the mobile view.''}

\begin{scriptsize}
\begin{verbatim}
--- Turn 1 ---
Agent thought: Mobile site editing falls
under "wix-platform-overview" which covers
editors, CMS, and design.

Tool: code_execution
  cat skill-01-wix-platform-overview/
    SKILL.md

Result: 14 subgroups. Agent picks
  group-13-wix-mobile-apps covering
  mobile editor features.

Tool: code_execution
  cat group-13-wix-mobile-apps/INDEX.md

Result: Leaf groups covering mobile UI
  elements. Agent finds group-00 with docs
  on Quick Action Bar (6 docs).

Tool: code_execution
  cat group-00-mobile-quick-action-bar/
    INDEX.md

Result: 6 documents including:
  - bce5e3d829fe: "Adding and Setting Up
    Your Quick Action Bar"

--- Turn 2 ---
Tool: get_document
  {"doc_id": "bce5e3d829fe6eeb"}

--- Answer ---
"There is no hide option, only deletion.
Right-click the Quick Action Bar in the
mobile Editor and select Delete. To re-add
later, go to Mobile Tools and click Add."

Navigation: 6221 -> 1770 -> 60 -> 6 -> 1
\end{verbatim}
\end{scriptsize}

\paragraph{Trace 4: Cross-branch exploration (Factuality 1.0, CtxR 1.0).}
Query: \emph{``I want to change the currency for my course.''}

\begin{scriptsize}
\begin{verbatim}
[Context: agent sees 6 skill descriptions]

--- Turn 1 ---
Agent thought: "Course" suggests Online
Programs under "wix-platform-ecosystem",
and "currency" is a payments topic under
"wix-commerce-operations".

Tool: code_execution
  cat skill-02-wix-platform-ecosystem/
    SKILL.md

Result: 11 subgroups listed, including:
  - group-07-wix-programs-partnerships/
    (119 docs): "online programs, courses,
    pricing, and partner integrations"

Tool: code_execution
  cat group-07-wix-programs-partnerships/
    INDEX.md

Result: 6 leaf groups. Picks group-04
  covering online program setup (9 docs).

Tool: code_execution
  cat group-04-wix-online-programs-setup/
    INDEX.md

Result: 9 documents including:
  - 3b2e9b0c4273: "Wix Online Programs:
    Setting Course Pricing"

Tool: get_document
  {"doc_id": "3b2e9b0c4273f8c9"}

Result: Course pricing article. Mentions
  that currency is configured at the site
  level via Language & Region settings,
  not per-program.

--- Turn 2 ---
Agent thought: Need the actual currency
configuration steps. Navigate to payments.

Tool: code_execution
  cat skill-00-wix-commerce-operations/
    SKILL.md | grep -i currency

Result: group-09-payment-provider-
  integration mentions currency settings.

Tool: code_execution
  cat group-09-payment-provider-integration/
    group-03-wix-currency-settings/INDEX.md

Result: 12 documents including:
  - 29139bd7c379: "Setting Your Currency
    for Accepting Payments"

Tool: get_document
  {"doc_id": "29139bd7c379ff77"}

Result: Step-by-step currency configuration
  guide via Settings > Language & Region.

--- Answer ---
"To change the currency for your course,
update the site-level currency setting:
1. Go to Settings in your dashboard.
2. Click Language & Region.
3. Scroll to Currency and select your
   desired currency.
4. Click Save.
This applies to all Wix products including
Online Programs. Your currency must match
your payment provider's currency."

Navigation paths:
  6221 -> 1797 -> 119 ->  9 -> 1 (programs)
  6221 -> 1682 -> 268 -> 12 -> 1 (currency)
\end{verbatim}
\end{scriptsize}

This trace demonstrates cross-branch navigation: the agent first explored the online-programs branch in one skill to understand the relationship between courses and currency, then cross-jumped to the payment-provider branch in a different skill to find the authoritative currency configuration steps.
The two branches across two skills provided complementary information that neither alone would have answered the query fully.

\section{Positioning Comparison}
\label{app:positioning}

\begin{table*}[h]
\centering
\small
\setlength{\tabcolsep}{5pt}
\renewcommand{\arraystretch}{1.15}
\caption{Positioning of \corpusskill{} relative to prior work.}
\label{tab:positioning}
\begin{tabular}{l ccccc}
\toprule
\textbf{Aspect} & \textbf{Scatter/Gather} & \textbf{RAPTOR} & \textbf{GraphRAG} & \textbf{Voyager/SkillX} & \textbf{\corpusskill{}} \\
\midrule
Input & Documents & Text chunks & Documents & Trajectories & Documents \\
Organization & Flat clusters & Hier. tree & Knowledge graph & Skill library & Hier. skill tree \\
Navigation & Human user & Algorithm & Graph search & Vector sim. & LLM agent \\
Serve infra & Cluster index & Vector index & Graph DB & Vector DB & \textbf{LLM only} \\
\bottomrule
\end{tabular}
\vspace{-0.5em}
\end{table*}

\Cref{tab:positioning} highlights three axes on which \corpusskill{} differs from prior work.

\paragraph{Input modality.}
Scatter/Gather, RAPTOR, GraphRAG, and \corpusskill{} all start from documents or text chunks, but Voyager and SkillX build skills from agent \emph{trajectories}: successful action sequences or distilled experience.
\corpusskill{} is the only system that converts a static document corpus into the skill format typically reserved for procedural knowledge.

\paragraph{Navigation mechanism.}
Scatter/Gather relies on human users to iteratively select clusters; RAPTOR and GraphRAG delegate navigation to fixed algorithms (vector similarity or graph traversal).
\corpusskill{} is unique in delegating navigation to an LLM agent that can reason about which branch to explore, backtrack when a path is unproductive, and combine evidence across branches.
This agent-driven navigation is what enables the trial-and-error behavior illustrated in \Cref{app:traces}.

\paragraph{Serve-time infrastructure.}
All prior systems require dedicated infrastructure at query time: a cluster index, a vector database, or a graph database.
\corpusskill{} eliminates this dependency entirely: the only runtime component is the LLM itself, which receives pre-compiled skill files via the API.
This ``LLM only'' property simplifies deployment but shifts the cost burden to input tokens, as discussed in \Cref{sec:limitations}.

\section{Failure Analysis}
\label{app:failure_analysis}

We examine the 45 of the 200 WixQA test queries (22\%) on which \corpusskill{} scores Factuality$\leq$0.4 or Context Recall$\leq$0.2, and categorize them into four failure modes.

\paragraph{Navigation miss (19 queries).}
The agent retrieves documents but none overlap with the gold context (Context Recall$=$0), indicating it descended into the wrong branch.
For example, a query about connecting PayPal is routed to a general payments group instead of the payment-setup leaf; a query about changing a header background on scroll navigates to a layout group rather than the header-effects cluster.
These misroutes typically occur at the \emph{top level}, where broad skill summaries fail to distinguish closely related topics.

\paragraph{Partial navigation (16 queries).}
The agent's retrieved set partially overlaps with the gold ($0<$Context Recall$<0.6$) but includes too many non-relevant documents, diluting the context and producing a low-quality answer.
The agent identified the correct general topic area but over-retrieved from neighboring leaf groups.

\paragraph{Synthesis error (5 queries).}
The agent retrieves contextually relevant documents (Context Recall$\geq$0.6) but produces a factually incorrect answer, typically by over-generalizing across multiple articles or misinterpreting conditional instructions as universal.

\paragraph{No documents retrieved (5 queries).}
Five queries resulted in zero or near-zero retrieval. These include off-domain queries (``How do I clear my browser cache?'', a Gmail password reset question, Google Ads billing) where the agent could not locate a matching skill branch and declined to answer rather than hallucinating.

\paragraph{Takeaway.}
The dominant bottleneck remains initial topic routing: 19 of 45 failures stem from the agent choosing the wrong top-level branch.
The narrow-tree ablation (\Cref{sec:ablations}), which produces finer-grained top-level clusters, partially mitigates this by providing sharper topic separation at the first navigation step.

\section{Baseline Implementation Details}
\label{app:baselines}

All baselines use the same LLM (Claude Sonnet) for answer generation and the same evaluation prompts for fair comparison.

\paragraph{BM25.}
We implement BM25 sparse keyword retrieval using the \texttt{rank\_bm25} library \citep{robertson2009bm25}.
Documents are tokenized and indexed at the full-article level.
At query time, we retrieve the top-5 articles by BM25 score and pass them as context to a single LLM call.

\paragraph{Dense.}
We encode all documents using Qwen3-Embedding-0.6B \citep{zhang2025qwen3embed}, producing 1024-dimensional embeddings stored in a FAISS flat index \citep{johnson2019faiss}.
We use the document prompt template for encoding articles and the query prompt template for encoding questions.
At query time, we retrieve the top-5 nearest neighbors by cosine similarity and pass them as context to a single LLM call.

\paragraph{Hybrid.}
We combine BM25 and Dense scores using Reciprocal Rank Fusion \citep{cormack2009rrf} with $k{=}60$ and equal weighting ($\alpha{=}0.5$).
The fused ranking selects the top-5 passages, which are passed to a single LLM call.

\paragraph{RAPTOR.}
We faithfully implement the RAPTOR algorithm \citep{sarthi2024raptor}: UMAP \citep{mcinnes2018umap} for dimensionality reduction, Gaussian Mixture Model (GMM) soft clustering \citep{mclachlan2000gmm} with BIC-based cluster count selection \citep{schwarz1978bic}, and recursive summarization of clusters.
The recursive cluster summarization uses Claude Sonnet~4.6, the same summarization model as the \corpusskill{} compiler, so the two embed-cluster-summarize pipelines are matched on the summarizer.
We use the same embedding model (Qwen3-Embedding-0.6B) as the Dense baseline for consistency; \Cref{app:stats} reports the encoder-controlled cross in which both \corpusskill{} and the baselines are run at either encoder size.
At query time, collapsed tree retrieval searches across all hierarchy levels simultaneously, returning the top-5 passages from any level.

\paragraph{Agentic.}
An LLM agent is given access to three search tools (\texttt{search\_bm25}, \texttt{search\_dense}, \texttt{search\_hybrid}), each returning top-$k$ results.
The system prompt instructs the agent to search first, then answer based on accumulated evidence, choosing BM25 for exact terms and product names, dense for conceptual queries, and hybrid as a default strategy.
The agent can issue up to 10 rounds of iterative retrieval, reformulating sub-queries based on intermediate results.
Evidence is accumulated across all turns; when the agent stops, it produces a structured JSON response containing the answer and curated top-5 references selected from the full evidence pool.
For evaluation fairness, context recall is computed over the last 5 retrieved documents (matching the context window of single-shot baselines).
This baseline tests whether agentic multi-turn retrieval over traditional search infrastructure can match navigation-based exploration.

\section{Evaluation Metrics}
\label{app:metrics}

\paragraph{Lexical and semantic answer-quality metrics.}
\emph{Token F1} tokenizes both the predicted and gold answers into word tokens and computes the harmonic mean of token-level precision and recall.
\emph{BERTScore-F1} \citep{zhang2020bertscore} computes contextual-embedding similarity F1 between the predicted and gold answers using a RoBERTa-large encoder \citep{liu2019roberta} with \texttt{rescale\_with\_baseline=False}; it is more robust to paraphrase than n-gram-overlap metrics. We compute it once per query and report the mean.

\paragraph{LLM-judged answer-quality and grounding metrics.}
\emph{Factuality}, \emph{Faithfulness$_{\text{grd}}$}, \emph{Context Recall}, \emph{Context Precision}, and \emph{Answer Relevance} use an LLM judge \citep{zheng2023llmjudge} following the RAGAS framework \citep{es2024ragas}, scoring on a 5-point scale normalized to 0--1.
Factuality compares the generated answer to the gold answer (independent of retrieved context).
Faithfulness$_{\text{grd}}$ scores each generated claim against the retrieved context, flagging hallucinations regardless of whether the answer happens to match the gold.
Context Recall asks whether the retrieved context covers the gold-answer claims; Context Precision asks whether retrieved chunks are predominantly relevant; Answer Relevance asks whether the generated answer addresses the query.
We use Claude Sonnet 4.6 as the judge with greedy decoding and zero-temperature prompts, and apply the same judge model and standardized context window across every method to keep the comparison fair.

\paragraph{Hallucination Rate.}
We report \emph{Hallucination Rate} as the fraction of queries with Faithfulness$_{\text{grd}}<0.6$ (i.e.\ generated content that is not adequately grounded in retrieved context, following the RAGAS convention).
This metric is lower-is-better and captures the operational ``how often does the system hallucinate?'' question that a deployer cares about, rather than a mean over a continuous score.

\paragraph{Context Recall standardization.}
A critical design decision is the standardization of Context Recall measurement across different retrieval paradigms (single-shot, multi-turn, and navigation-based).
For fair comparison, all methods follow the same protocol:
(1)~collect the last 5 documents or passages retrieved by the method;
(2)~join them with separator tokens;
(3)~truncate to a uniform 8,000-character window.
The LLM judge receives this standardized context.
This ensures that single-shot baselines contribute their top-5 retrieved passages, Agentic contributes the last 5 documents from its multi-turn retrieval, and \corpusskill{} contributes the last 5 documents retrieved via \texttt{get\_document} tool calls.

\paragraph{Cost metrics.}
Input token count is the total number of tokens sent to the LLM across all turns.
Per-query cost is computed using the model's published per-token pricing (input and output tokens combined) including the Anthropic ephemeral-cache discount (cached-read input at 0.1$\times$ the base rate, cache-write at 1.25$\times$).
For single-shot methods, cost reflects one LLM call; for multi-turn methods (Agentic, \corpusskill{}), it reflects all turns including tool-call reasoning.

\section{Dataset Licenses, Intended Use, and Documentation}
\label{app:licenses}

\paragraph{Licenses.}
We use two publicly released benchmark datasets.
\emph{WixQA}~\citep{cohen2025wixqa} is distributed on Hugging Face\footnote{\url{https://huggingface.co/datasets/Wix/WixQA}} under the \textbf{MIT License}, which permits unrestricted use, modification, and redistribution with attribution.
\emph{RAGBench}~\citep{friel2024ragbench} is distributed on Hugging Face\footnote{\url{https://huggingface.co/datasets/galileo-ai/ragbench}} under the \textbf{Creative Commons Attribution 4.0 International (CC-BY-4.0)} license, which permits sharing and adaptation for any purpose provided appropriate credit is given.
Our own code is released at \codeurl{} under the MIT License.

\paragraph{Consistency with intended use.}
Both datasets were created as benchmarks for evaluating retrieval-augmented generation systems, which is exactly how we use them.
WixQA was designed to evaluate enterprise RAG pipelines over a customer-support knowledge base; we use its 200 expert-written test queries and 6{,}221-article corpus for our primary evaluation (\Cref{sec:experiments}).
RAGBench was designed as a multi-domain RAG evaluation benchmark; we use 10 of its 12 subsets for our generalization study (\Cref{sec:ragbench}).
We do not redistribute modified versions of either dataset.
Our derived artifacts (compiled skill hierarchies) are deterministic transformations of the original corpora and are intended solely for research evaluation.

\paragraph{Documentation and coverage.}
All experiments use English-language data exclusively.
WixQA covers the enterprise customer-support domain (Wix.com Help Center articles spanning commerce, site design, billing, SEO, and platform management) and contains 6{,}221 documents with three QA subsets: 200 expert-written queries, 200 simulated queries, and 6{,}221 synthetic queries.
The 10 RAGBench subsets span five broad domains: medical (CovidQA), legal (CUAD), technical support (TechQA, EManual, DelucionQA), financial (FinQA, TatQA), and open-domain (HAGRID, ExpertQA, ClapNQ), with corpus sizes ranging from 221 to 12{,}727 documents per subset.
Neither dataset contains personally identifiable information; WixQA consists of public help-center articles and RAGBench aggregates previously published academic datasets.
We do not make claims about demographic representation, as both benchmarks are topical QA datasets rather than user-facing demographic studies.
Full details on corpus statistics and evaluation protocols are provided in \Cref{sec:experiments} and \Cref{app:baselines}.

\section{Statistical Testing and Additional Controls}
\label{app:stats}

This appendix gives the statistical protocol behind every win/tie/loss statement and the full results of the controls summarized in \Cref{sec:ablations} and \Cref{sec:scale}. Unless stated otherwise, tables use the primary setup: WixQA (6{,}221 support articles, 200 expert-written queries), Claude Sonnet~4.6 as compile and serving LLM, Qwen3-Embedding-8B as compile encoder, and Token F1 against the gold answer as the metric.

\paragraph{Paired bootstrap protocol.}
For each dataset we pair \corpusskill{}'s per-query Token F1 with the best baseline's on the matched queries and resample the paired differences 10{,}000 times.
A dataset is a \emph{win} if the two-sided $p$-value for a positive mean difference is below 0.05, a \emph{loss} in the reverse direction, and a \emph{tie} otherwise.
Per-metric confidence intervals are in the released code, and the same procedure underlies the deltas in \Cref{tab:app_raptor_delta}.

\begin{table}[!htbp]
\centering
\scriptsize
\setlength{\tabcolsep}{4pt}
\caption{Encoder cross on WixQA (Token F1, 200 queries per cell): the encoder choice barely moves \corpusskill{} and helps the baselines by at most $+$0.023.}
\label{tab:app_encoder_cross}
\begin{tabular}{l cc}
\toprule
\textbf{Method} & 0.6B encoder & 8B encoder \\
\midrule
Dense  & 0.364 & 0.378 \\
Hybrid & 0.361 & 0.360 \\
RAPTOR & 0.398 & 0.421 \\
\corpusskill{} & 0.461 & 0.458 $\pm$ 0.002 (3 runs) \\
\bottomrule
\end{tabular}
\end{table}

\begin{table}[!htbp]
\centering
\scriptsize
\setlength{\tabcolsep}{3pt}
\caption{\corpusskill{} $-$ RAPTOR, paired bootstrap on WixQA (200 queries, 10k resamples), with compile resources matched at either encoder size. RAPTOR uses the same Sonnet~4.6 summarizer in all rows.}
\label{tab:app_raptor_delta}
\begin{tabular}{l cc}
\toprule
\textbf{Metric} & matched at 0.6B & matched at 8B \\
\midrule
Token F1          & $+$0.064 ($p{<}0.0001$) & $+$0.035 ($p{<}0.0001$) \\
BERTScore         & $+$0.025 ($p{<}0.0001$) & $+$0.020 ($p{<}0.0001$) \\
Context Precision & $+$0.181 ($p{<}0.0001$) & $+$0.085 ($p{<}0.0001$) \\
Context Recall    & $+$0.084 ($p{=}0.007$)  & $-$0.036 ($p{=}0.22$, tie) \\
\bottomrule
\end{tabular}
\end{table}

\begin{table}[!htbp]
\centering
\scriptsize
\setlength{\tabcolsep}{2.5pt}
\caption{Per-dataset Token F1 verdict (paired bootstrap vs.\ the best baseline per dataset). $\Delta$ is the mean per-query paired difference over matched queries and can differ slightly from the difference of column means.}
\label{tab:app_verdicts}
\begin{tabular}{l l c c r l l}
\toprule
\textbf{Dataset} & \textbf{Best bl.} & C2S F1 & Bl.\ F1 & $\Delta$ & $p$ & \textbf{Verdict} \\
\midrule
WixQA      & RAPTOR  & 0.456 & 0.397 & $+$0.059 & $<$0.001 & win \\
Emanual    & Agentic & 0.497 & 0.455 & $+$0.061 & $<$0.001 & win \\
DelucionQA & Dense   & 0.478 & 0.439 & $+$0.034 & 0.010    & win \\
ExpertQA   & Dense   & 0.356 & 0.330 & $+$0.027 & $<$0.001 & win \\
ClapNQ     & Dense   & 0.468 & 0.333 & $+$0.135 & $<$0.001 & win \\
\midrule
CovidQA    & Hybrid  & 0.208 & 0.216 & $-$0.009 & 0.33     & tie \\
TechQA     & Hybrid  & 0.396 & 0.383 & $+$0.013 & 0.14     & tie \\
FinQA      & Dense   & 0.392 & 0.400 & $-$0.009 & 0.49     & tie \\
\midrule
CUAD       & BM25    & 0.272 & 0.314 & $-$0.052 & $<$0.001 & loss \\
HAGRID     & Agentic & 0.188 & 0.356 & $-$0.167 & $<$0.001 & loss \\
TatQA      & Agentic & 0.240 & 0.355 & $-$0.116 & $<$0.001 & loss \\
\bottomrule
\end{tabular}
\end{table}

\paragraph{Serve-time variance.}
Three independent 200-query serve passes over the same compiled WixQA tree give (mean $\pm$ sd) Token F1 0.458 $\pm$ 0.002, BERTScore 0.863 $\pm$ 0.001, Factuality 0.767 $\pm$ 0.008, Context Recall 0.699 $\pm$ 0.010, and Hallucination Rate 0.065 $\pm$ 0.018. The thresholded hallucination rate is the most variable metric.

\paragraph{Cost without prompt caching.}
Removing the ephemeral-cache discount from the raw token logs raises \corpusskill{} from \$0.153 to \$0.302 per query ($1.97\times$) and Agentic from \$0.082 to \$0.098 ($1.19\times$); the ordering is unchanged.

\end{document}